\documentclass[epj,final]{svjour}
\usepackage{subfigure}       % permite a inclus\~ao de subfiguras
\usepackage{graphicx}     % visuaiza apenas a regi\~ao ocupada pela figura EPS
\usepackage[sort&compress]{natbib}
\usepackage{hyperref}

\begin{document}
\title{Critical vortex line length near a zigzag of pinning centers}
\author{Antonio R. de C. Romaguera\thanks{ton@if.ufrj.br}\and Mauro M. Doria\thanks{mmd@if.ufrj.br}% etc
}
\institute{ Instituto de F\'{\i}sica, Universidade Federal do Rio
de Janeiro
\\C. P. 68528 CEP 21941-972, Rio de Janeiro, RJ, Brazil
}
%
%\date{Received: April 2004 / Revised version: July 2004}
%
\abstract{ A vortex line  passes through as many pinning centers
as possible on its way from one extremety of the superconductor to
the other at the expense of increasing its self-energy. In the
framework of the Ginzburg-Landau theory we study the relative
growth in length, with respect to the straight line, of a vortex
near a zigzag of defects. The defects are insulating pinning
spheres that form a three-dimensional cubic array embedded in the
superconductor. We determine the depinning transition beyond which
the vortex line no longer follows the critical zigzag path of
defects.
\PACS{ {74.80.-g}{Spatially inhomogeneous structures}\and
{74.25.-q}{General properties; correlations between physical
properties in normal and superconducting states}\and
{74.20.De}{Phenomenological theories (two-fluid, Ginzburg-Landau,
etc.)}
     } % end of PACS codes
\keywords{{Ginzburg-Landau} \and {Tridimensional} \and {pinning spheres}}%
} %end of abstract
\maketitle
\section{Introduction}
\label{sec:int}

The magnetic field penetrates the superconductor in the form of
filaments that pierce the sample from one extremety to the other.
The filament is a vortex line carrying a quantized unit of
magnetic flux, $\Phi_0=hc/2e$, where in its core the magnetic
field reaches its maximum value inside the superconductor. The
vortex line is not rigid and many factors can contribute to its
length such as thermal fluctuations, the geometry of the sample,
anisotropy, and most important of all, imperfections inside the
superconductor, specially pinning centers. At the center of a flat
sample, for instance, with the applied magnetic field
perpendicular to the main surface, the length of the vortex line
is the height of the sample. However the vortex line will tend to
adjust its length to the distribution of pinning centers so to
pass through as many of them as possible resulting in a  vortex
line bigger than the height of the sample. The nucleation of a
vortex line on a pinning center is advantageous\cite{MS72} for the
superconductor because both the core of the vortex line and the
pinning center are non-superconducting regions. Nevertheless long
detours optimize the number of pinning centers visited by the
vortex to a maximum, but this  makes the vortex line too long and
this increases its self-energy, which grows with its length.

The study of pinning centers is fundamental to the understanding of superconductors\cite{B95}.
The interaction of pinning centers with vortices has been studied using several approaches \cite{KetSongbook,B95}.
Superconducting samples are plagued with natural pinning centers and a way to attack the problem is to consider samples with artificially made pinning centers\cite{PM04}, such as columnar defects\cite{Bu93,LGTBHP96}, antidots\cite{MBMRBTBJ98,YLBJW04} and micro holes\cite{BP95}.
They are interesting because they bring clear-cut questions about the interaction between vortices
and pinning centers, such as the one we are interested here concerning the local misalignment of a vortex line.
A vortex line is aligned to the magnetic induction direction in the absence of pinning centers, but in presence of them
bends  and acquires a new shape, though it remains globally oriented along the magnetic induction.

In the classical problem of the travelling salesman
\cite{Numrecbook} the seller must find the shortest route that
connects several cities at well known locations that he must
visit. This is a minimization problem that in case of a large
number of cities has many local minima, that is, many possible
routes  very close in length to the shortest one. Similarly to the
travelling salesman, one may wonder what is the maximum length
that a vortex line can reach inside a superconductor with pinning
centers. To study the relative maximum length we consider the
vortex line near a zigzag of pinning centers, namely, a path with
abrupt alternate right and left turns such that at each turn there
is a pinning center. The pinning centers are just of one kind,
insulating spheres of radius $R$. Figures \ref{figure1} and
\ref{figure2} give a pictorial view of this superconductor with
pinning centers. The vortex line acquires this zigzag shape as
long as trapping by the pinning centers is advantageous as
compared to the increase in length caused by the zigzag path.
There is a critical path that sets a depinning transition beyond
which the vortex line does not follow the zigzag path of the
pinning centers. Here we numerically determine this transition
without considering thermal fluctuations. We treat the problem in
the framework of the Ginzburg-Landau theory\cite{A57}. From the
point of view of the Ginzburg-Landau theory, pinning may be caused
by spatial fluctuations of the critical temperature, $T_c(\vec
x)$\cite{L70}, or of the mean free-path that changes the
coefficient in front of the gradient term,  $\xi(\vec x)^2|({\vec
\nabla} - {{2\pi i}\over{\Phi_0}} {\vec A})\Delta|^2$. The
interaction between a vortex line and a pinning center has been
considered by many authors in the context of the Ginzburg-Landau
theory\cite{DA99,DZ02a,DZ02b,PF03}.

The paper is organized as follows. In  Section ~\ref{sec:mod} we
describe the present model of a superconductor with pinning centers,
and in Section ~\ref{sec:theo} our theoretical approach is discussed.
In Section ~\ref{sec:res} we give our results
obtained through numerical simulations. In Section ~\ref{sec:con}
we summarize the main achievements of this work.

%......................................................
\section{Model}
\label{sec:mod}
\begin{figure}[b]
\centering
\includegraphics[width=0.8\linewidth]{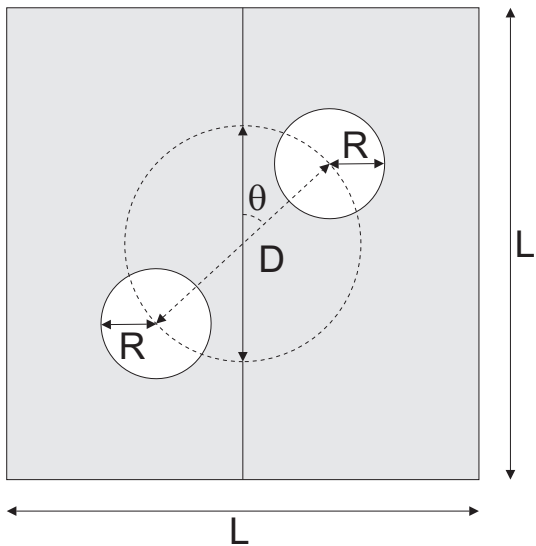}
\caption{Position of the defects inside the unit cell. The
superconducting material occupies the filled region.}
\label{figure1}
\end{figure}

\begin{figure}[t]
\centering
\includegraphics[width=0.7\linewidth]{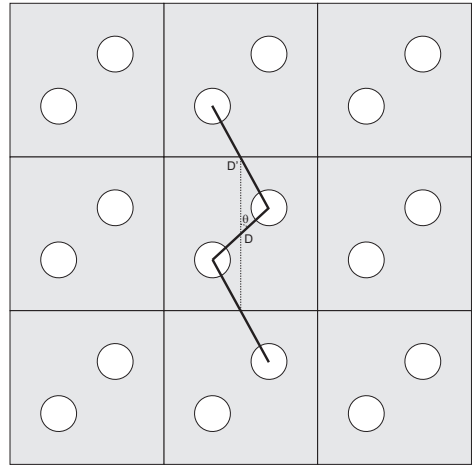}
\caption{Pictorial view of several unit cells forming the pinning
center lattice.} \label{figure2}
\end{figure}

The model consists of a superconductor with a cubic array of
defects, described by its simplest unit cell, a cube of size $L$
containing two pinning spheres inside, separated by $D$, as shown
in figure \ref{figure1}.
%---------------------figure  #1------------------------------------------------------
%---------------------figure  #1------------------------------------------------------
The geometric center of these two pinning spheres coincides with
the center of the unit cell. The general pinning center
distribution is obtained by rotating the center of the segment $D$
by an angle $\theta$ with respect to the axis along which is
oriented the magnetic induction, hereafter called z-axis, on a
fixed plane defined by both x and z axes. The magnetic flux that
crosses the face perpendicular to the z axis of the unit cell is
$\Phi_0$ and each vortex line is near a zigzag of pinning spheres,
made by the pile of unit cells along the z-axis, according to
figure \ref{figure2}.
%---------------------figure  #2------------------------------------------------------
%---------------------figure  #2------------------------------------------------------

Though the present model treats a vortex lattice, the depinning
transition is that of a single isolated vortex line near the
zigzag of defects. The vortex-vortex interaction has no effect on
this transition since the vortex-defect interaction is short
ranged and very strong. In this paper we obtain the depinning
transition for the cubic lattice of defects with $L = 12.0 \xi$,
$D = 6.0 \xi$, and several pinning sphere radii are considered,
ranging from $R=1.2 \xi$ to the maximum possible radius, $3.0
\xi$, with an increment of $0.20\xi$.

To understand the length of the vortex line in this model, first
consider $\theta=0^\circ$, where  the zigzag collapses into a
straight line made of two alternate segments, $D$ and $L-D$. Thus
the maximum pinning sphere that fits this unit cell has radius
equal to $(L-D)/2$. For arbitrary $\theta$, the segment connecting
the top (bottom) pinning center to the nearest neighbor pinning
center in the unit cell above (below), is
$D'=\sqrt{(L-D)^2+4LD\sin^2{(\theta/2)}}$. Notice that the pinning
center density is $\theta$ independent, and equal to $2/L^3$. The
depinning transition is described either by the critical angle
$\theta_c$, or by the relative pinning length, $\Delta l/l\big|_c
= [l(\theta_c)-l(0^\circ)]/l(0^\circ)$, $l(\theta_c)$ being the
critical length of the zig-zag path and $l(0^\circ)$ the length of
the straight line formed by alternate $D$ and $D'$ segments along
the z-axis. From geometry one obtains that,
\begin{equation}
\frac{\Delta l }{l}(\theta)
=\frac{\sqrt{\left(L-D\right)^2+4DL\sin^2{(\theta/2)}}-(L-D)}{L}
\label{deltal}
\end{equation}
It is more interesting to characterize the depinning transition
through the relative pinning length $\Delta l/l$ because it also
describes, though approximately, the relative length growth of the
vortex line  due to the presence of the zigzag of pinning centers.
The maximum zigzag path possible, and consequently, the maximum
length that the vortex line can achieve, occurs for
$\theta=90^\circ$. Thus for the present set of parameters the
maximum relative growth of the vortex line possible is $(\Delta
l/l)_{max} = (\sqrt{5}-1)/2$, that is the line can be $61,8 \%$
bigger than a straight line, as found from Eq.(\ref{deltal}).
%
%............................................................................
\section{Theoretical approach}
\label{sec:theo}
The numerical search for the free energy minima is carried for the
unit cell with the two pinning spheres inside, taking care that
the well-known Saint-James de Gennes \cite{deGennesbook} boundary
condition is satisfied at their surfaces because they are
insulating spheres. The radial component of the supercurrent must
vanish there. The unit cell has quasi-periodic boundary conditions
so to describe a cubic array of pinning spheres embedded in the
superconductor. This system is studied here using a
Ginzburg-Landau free energy expansion that considers
superconducting and non-superconducting regions on equal footing
and enforces  this boundary condition during the
minimization procedure.

Notice that the choice of the simplest unit cell helps the
minimization concerning the computational time, but from the other
side forbids the onset of vortex line configurations other than
the zigzag one. More complex vortex configurations in that cubic
lattice of pinning centers require a larger unit cell to be
studied. The present numerical procedure uses a mesh of $P^3$
points to describe the unit cell, and for the present simulations
we take that $P=19$. Thus the distance between two consecutive
mesh points is $a=2\xi/3$, and one obtains that $L=2\xi(P-1)/3$.
The local field $\vec h=\nabla \times\vec A$ is constant inside
the unit cell and equal to the magnetic induction, $\vec B$. The
regime treated here is of strong type-II superconductors with the
London penetration length much larger than the unit cell
($\lambda\gg\ L > \xi$). The free energy depends on $P^3$ times
the number of variables, and in our case, just the variables
$Re(\Delta)$ and $Im(\Delta)$ participate in the minimization
procedure. Under this considerations, we numerically minimized
$2P^3$ variables through the Simulated Annealing method, a Monte
Carlo thermal procedure. We take 1600 visits per site for each
Monte Carlo temperature, with 150 temperature reductions, though
this last number depends on how fast convergence to the absolute
minimum is achieved.

To obtain the vortex solution in presence of the two pinning
spheres, we minimize the Ginzburg-Landau free energy inside the
unit cell assuming that the magnetic flux that crosses the cube
along z perpendicular faces is  $\Phi_0$, and this corresponds to
a magnetic induction $\vec B = 2 \pi \kappa (\xi/L)^2 \hat z$, in
units $\sqrt{2}H_c$, where $\kappa$ is the Ginzburg-Landau
constant and  $H_c$ is the superconductor critical field. Under
this condition a single vortex nucleates inside the unit cell. In
a coordinate system whose origin is at the center of the unit
cell, and the axis are along the cube's main directions, the
positions of the pinning spheres, are $\vec x_1= R(\sin \theta
\hat x + \cos \theta \hat z)$, and $\vec x_2 = -\vec x_1$, as
specified in figure \ref{figure1}. In this approach the pinning
centers are described by  step-like functions, zero inside and one
outside, which are made smooth for numerical reasons, $\tau(\vec
x)= \tau_1(\vec x)\tau_2(\vec x)$, $\tau_i({\vec x}) = 1 - 2 /
\lbrace \; 1 + \exp{{\lbrack(|\vec x - \vec x_i|/R)}^{K}
\rbrack}\;\rbrace $, with $K=8$, $i=1,2$. The free energy density
is,
\begin{eqnarray}
\mathcal{F} = \int {{dv}\over{V}}&& \Biggl \{\;\tau(\vec x) \left[
\xi^2 |({\vec \nabla} - {{2\pi i}\over{\Phi_0}} {\vec A})\Delta|^2 -
|\Delta|^2 \right] \nonumber \\&& +  {1 \over 2} |\Delta|^4\Biggl
\}, \label{eq:glth}
\end{eqnarray}
expressed in units of the critical field energy density,
$H_c^2/4\pi$, and the superconducting density $|\Delta|^2$
normalized between zero and one. This free energy density takes
value in the range 0  and -0.5, its minimum and maximum,
respectively. These extremes correspond to the normal and the
spatially constant superconducting density states, respectively.
The constant density state differs from the normal state by a
constant, equal to the condensation energy $H_c^2/8\pi$. The
magnetic energy density also yields a constant taken to vanish for
each value of the magnetic induction. Because $\tau=0$ inside the
spheres, Eq.(\ref{eq:glth}) has the trivial solution $\Delta = 0$
with the condition $\hat n \cdot ({\vec \nabla} - {{2\pi
i}\over{\Phi_0}} {\vec A})\Delta\Big|_{surface} = 0$ satisfied at
the pinning spheres surfaces.

To gain some insight into the free energy expansion of
Eq.(\ref{eq:glth}), we look at the contribution that an empty
pinning sphere brings to the free energy in case of no vortices in
the unit cell. A defect-free superconductor reaches the maximum
density everywhere, $|\Delta|^2=1$,  and its free energy density
is  simply $\mathcal{F}_0=-0.5$, according to Eq.(\ref{eq:glth}).
The superconductor with the two pinning centers per unit cell has
a higher energy because inside the pinning centers the order
parameter should vanish, $|\Delta|^2=0$, rendering its free energy
density approximately  equal to the defect-free case removed of
the volume of the two spheres\cite{DR04}, thus equal to $\mathcal{F}_0(1-8\pi
R^3/3L^3)$. This result is only approximately valid since the
curvature of the order parameter near the pinning sphere surface
causes an increase in the kinetic energy, an effect that becomes
more pronounced for large spheres. The kinetic energy density,
\begin{equation}
\mathcal{F}_{kin} = \int {{dv}\over{V}} \; \tau \; \xi^2  |({\vec
\nabla} - {{2\pi i}\over{\Phi_0}} {\vec A})\Delta|^2, \label{fkin}
\end{equation}
is more sensitive to the depinning transition \cite{DSOGB02} than
the total free energy density. The term $\vec \nabla \Delta$ picks
a large contribution when the vortex detaches from a pinning
sphere  because the order parameter must bend at an extra
superconductor-insulator interface.
%
%
%.....................................................................................................................
\section{Results}
\label{sec:res}
\begin{figure*}[!ht]
\centering
    \subfigure[]
    {
    \begin{minipage}[t]{0.46\textwidth}
    \centering
    \includegraphics[width=\linewidth]{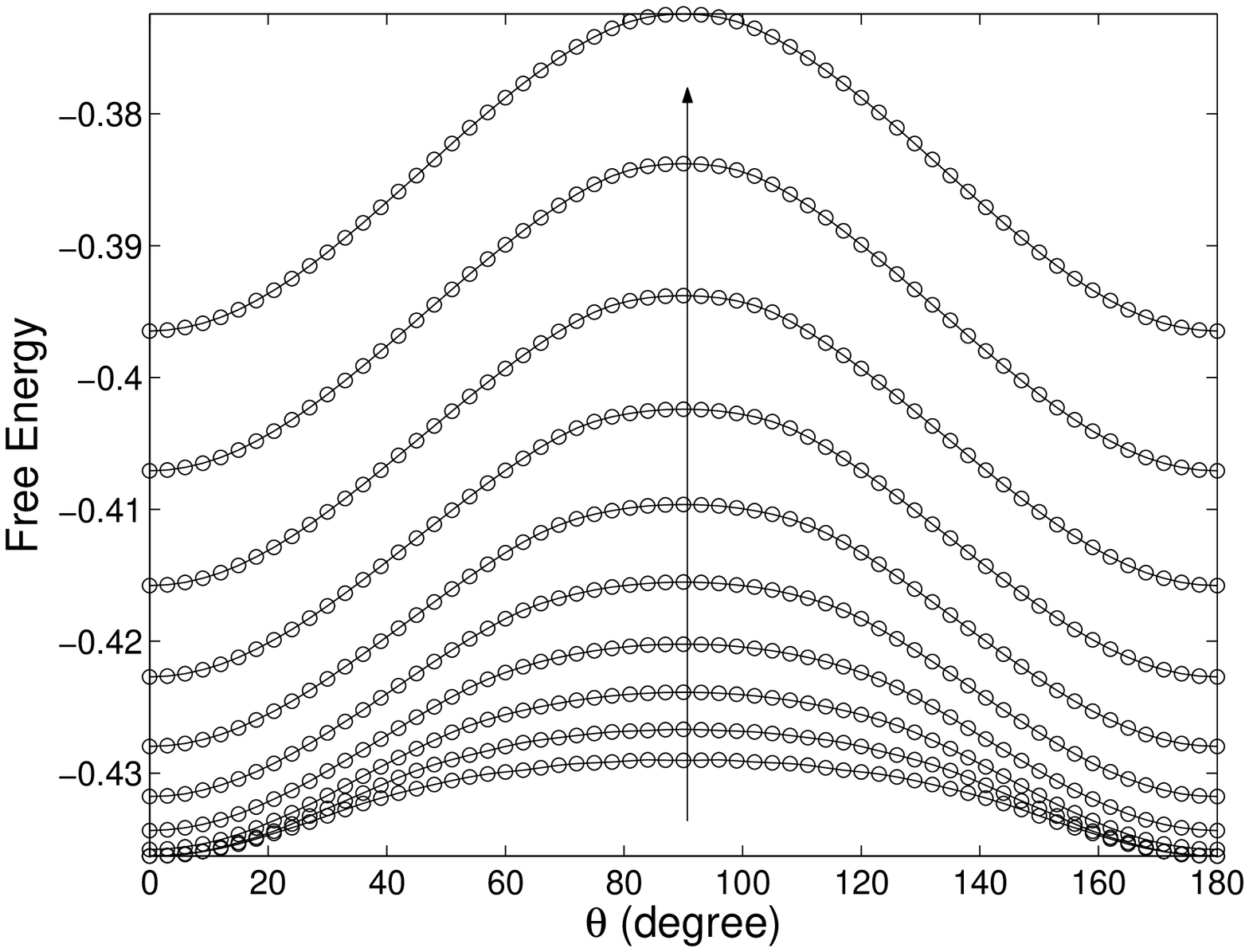}
    \label{figure3a}
    \end{minipage}
    }
    \subfigure[]
    {
    \begin{minipage}[t]{0.46\textwidth}
    \centering
    \includegraphics[width=\linewidth]{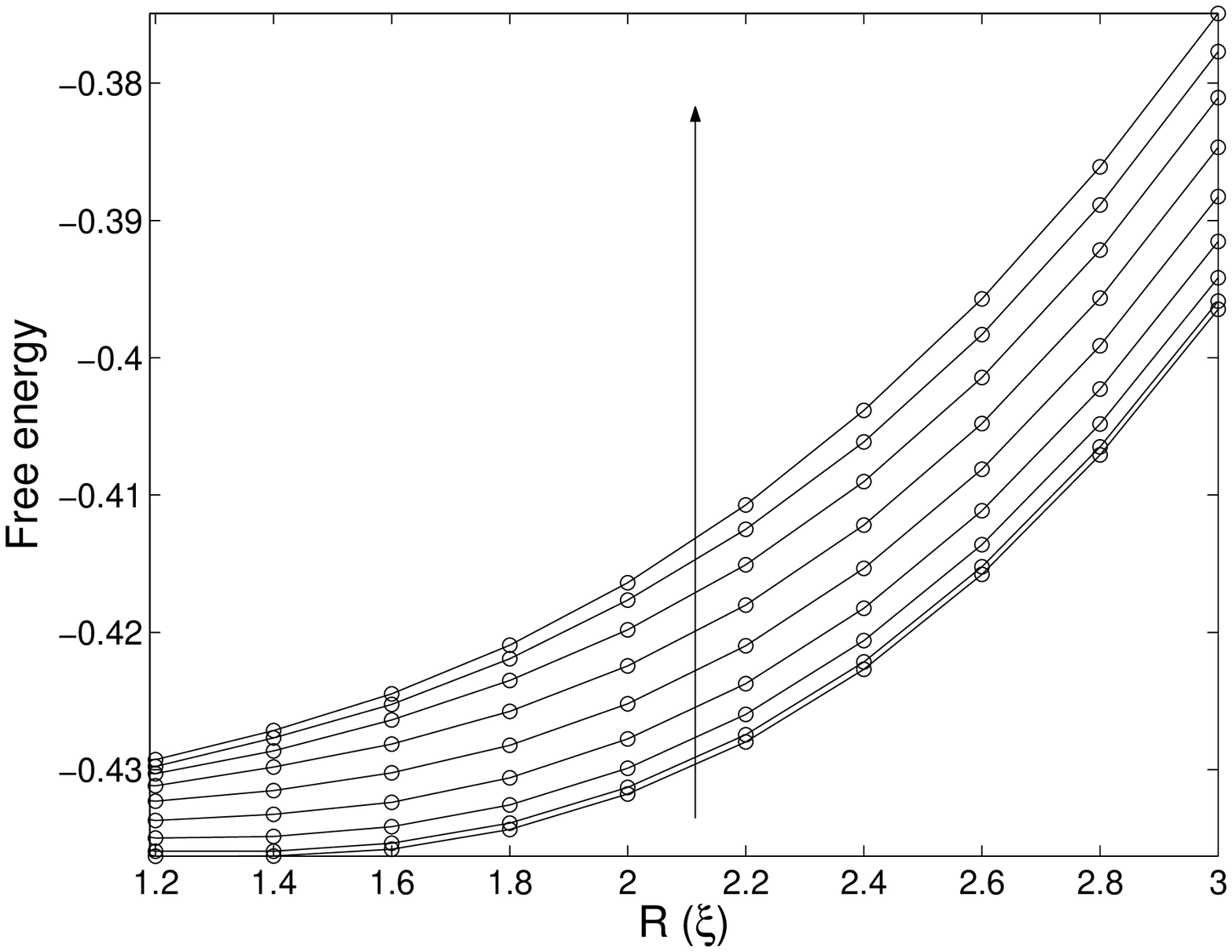}
    \label{figure3b}
    \end{minipage}
    }
    \caption
    {
    Dependency of the free energy, $\mathcal{F}$, with the pinning center radius $R$ and the
    angle $\theta$.
    \ref{figure3a}
    {Variation of
    $\mathcal{F}$ as a function of the angle $\theta$ in the range $0^\circ$ to $180^\circ$,
    data points obtained for increments of $3^\circ$. The radius $R$ varies from $1.2\xi$
    to $3.0\xi$ and for each increment of $0.2\xi$ results in a distinct curve,
    all plotted in ascendant order as specified by the arrow}.
    \ref{figure3b}
    {Variation of
    $\mathcal{F}$ as a function of the pinning center radius for a specific value of $\theta$.
    The radius varies from $1.2\xi$ to $3.0\xi$ with a increment of $0.2\xi$. Distinct curve correspond to different
    $\theta$, equal to $0^\circ$, $9^\circ$, $18^\circ$, $27^\circ$, $36^\circ$, $45^\circ$, $54^\circ$,
    $63^\circ$ and $72^\circ$ in the ascendant order indicated by the arrow.}
    }
    \label{figure3}
\end{figure*}

The free energy density versus $R$ is shown in figure
\ref{figure3a} from $0^\circ$ to $90^\circ$, in steps of
$9^\circ$. To understand the growth of the free energy with
respect to $R$, firstly consider, the $\theta=0^\circ$ curve, the
lowest one in free energy. The vortex line is pinned to the two
aligned  spheres along the z-axis. Considering the vortex core as
a non-superconducting cylinder of radius $\xi$, it follows that
for $R \le \xi$ the two pinning spheres are fully inside the
vortex core, but not for $R > \xi$. This extra volume, of the
insulating region outside the vortex core, grows with the size of
the spheres and, as previously discussed, makes the energy
approach zero, that is, the normal state, because there is less
superconducting volume in the unit cell. This behavior is seen in
figure \ref{figure3a}for any angle and not only for $0^\circ$. The
free energy density, as calculated by Eq.(\ref{eq:glth}), versus
$\theta$ is shown in figure \ref{figure3b} for several radii,
ranging from $R=1.2\xi$ to $R=3.0\xi$, in steps of $0.2\xi$. It
shows symmetry with respect to $90^\circ$, as expected, since the
two zigzag paths, associated to $\theta$ and to $180^\circ -
\theta$, are just mirror images of each other, and so, have the
same energy. For all radii,  the configuration of minimum energy
is for $0^\circ$ and the maximum for $90^\circ$, because the
$90^\circ$ arrangement has a smaller fraction of superconducting
volume  as compared to  the $0^\circ$ configuration. For $0^\circ$
the two spheres are aligned along the z-axis and both pin the
vortex line whereas for $90^\circ$ only one sphere pins the vortex
and the other one is free in space, and so there is less
condensate energy. This makes the superconductor closer in energy
to the normal state, thus increasing its energy according to
Eq.(\ref{eq:glth}). At some intermediate angle between $0^\circ$
and $90^\circ$ occurs the depinning transition, although  it is
not noticeable in  both figures  \ref{figure3a} and
\ref{figure3b}.
%---------------------figure  #3------------------------------------------------------
%---------------------figure  #3------------------------------------------------------
The depinning transition is not perceptible in
$\mathcal{F}(\theta)$, nor in its derivative with respect to
$\theta$, because of numerical
limitations, possibly due to the coarseness of the mesh used in our calculations. In summary figures
\ref{figure3a} and \ref{figure3b} provide different views of the
same data, which shows that the free energy density increases with
$R$, a property previously explained in terms of the superposition
of the vortex line with the pinning spheres.

The depinning transition is revealed by the kinetic energy of
Eq(\ref{fkin}), though it steadily  grows with $\theta$,
reaching its maximum and minimum at the extremes $0^\circ$ and
$90^\circ$, respectively.

%---------------------figure  #4------------------------------------------------------
\begin{figure}[!ht]
    \subfigure[]
    {
    \begin{minipage}[b]{0.9\linewidth}
    \includegraphics[width=\linewidth]{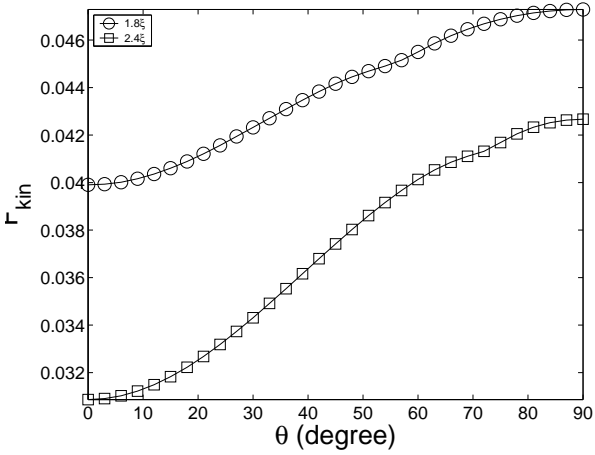}
    \label{figure4a}
    \end{minipage}
    }
    \subfigure[]
    {
    \begin{minipage}[b]{0.9\linewidth}
    \includegraphics[width=\linewidth]{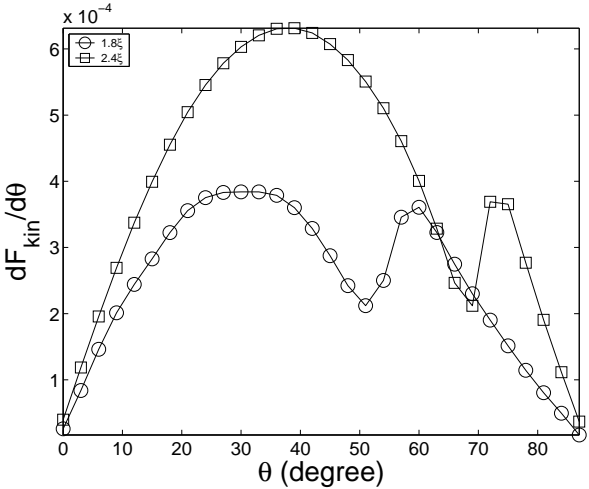}
    \label{figure4b}
    \end{minipage}
    }
    \caption{The kinetic energy and its derivative. In \ref{figure4a} the curves $F_{kin}\times \theta$
            for $R=1.8\xi$ and $R=2.4\xi$. In \ref{figure4b} their derivatives.}
    \label{figure4}
\end{figure}
%---------------------figure  #4------------------------------------------------------
The depinning transition is best seen in the  derivative
$d\mathcal{F}_{kin}/d\theta$ versus $\theta$ curve. Figures
\ref{figure4a} and \ref{figure4b} show the kinetic energy and its
derivative  for two selected radii, namely $R=1.8\xi$, and
$R=2.4\xi$. Both cases show that the kinetic energy derivative
vanishes for $0^\circ$ and $90^\circ$, within numerical precision,
which implies that the kinetic energy derivative must have at
least one maximum in between these extreme angles\cite{DSOGB02}.
Indeed these curves, as shown in figures \ref{figure4a} and
\ref{figure4b}, display a double hump structure with the lowest
angle one as the absolute maximum. The depinning transition
corresponds to the critical angle which defines the local minimum,
located between the two humps. The critical angle defined by this
procedure is listed in Table~\ref{tab:thetac} for several radii.
The reason for the second hump relies on the superposition of the
vortex core with the two pinning centers.
As long as it happens the order parameter has to adjust
around one single common interface. However above the depinning
transition, the vortex unpins from one of the spheres, yielding
two independent surfaces where the deflection of the order
parameter must take place. This leads to an extra growth of the
kinetic energy above the transition because an order parameter
gradient is set around  two surfaces instead of just one.

%---------------------table  #1------------------------------------------------------
\begin{table}[t]
\caption{The critical angle $\theta_c$ is given here, defined by
the local minimum of the curves of figures \ref{figure4}.}
\centering
\begin{tabular}{ll}
\hline
\noalign{\smallskip}
Radius ($\xi$) & $\theta_c (degree)$ \\
\noalign{\smallskip}\hline\noalign{\smallskip}
1.2 & 33$^\circ$ \\
1.4 & 42$^\circ$ \\
1.6 & 48$^\circ$ \\
1.8 & 54$^\circ$ \\
2.0 & 60$^\circ$ \\
2.2 & 66$^\circ$ \\
2.4 & 72$^\circ$ \\
2.6 & 75$^\circ$ \\
2.8 & 78$^\circ$ \\
3.0 & 81$^\circ$ \\
\noalign{\smallskip}
\hline
\end{tabular}
\label{tab:thetac}
\end{table}
%---------------------table  #1------------------------------------------------------

\begin{figure}[b]
\centering
\includegraphics[width=1.0\linewidth]{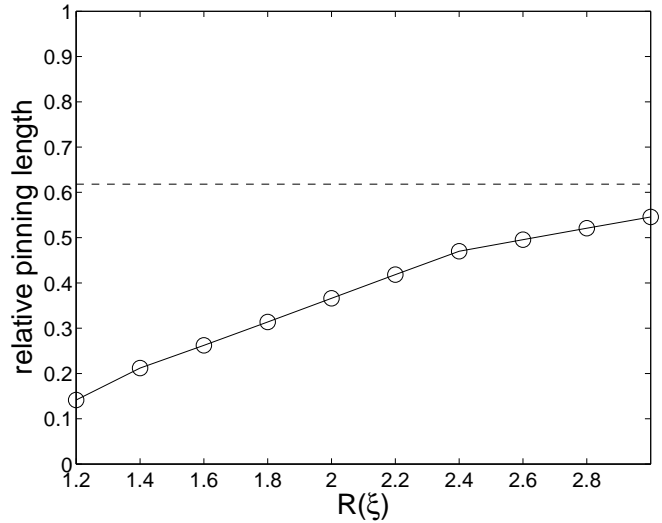}
\caption{Critical vortex line versus the radius of the pinning
sphere. The dashed line means the maximum relative pinning length
that is independent of the pinning center radius $R$.}
\label{figure5}
\end{figure}

The most relevant results of this paper are summarized in figure
\ref{figure5}, which shows the critical relative pinning length,
as computed from Eq.(\ref{deltal}) for several pinning sphere
radii. The content of this figure is that of
Table \ref{tab:thetac}. The usefulness of plotting $\Delta
l/l(\theta_c)$ instead of $\theta_c$ is to get direct information
about the relative increase of the vortex line in presence of the
zigzag arrangement of pinning centers. For instance, for the
selected radii, $R=1.8\xi$ and $R=2.4\xi$, figure \ref{figure5}
directly gives that the vortex line can stretch  to a maximum of
$31\%$ and $46\%$, respectively, in presence of this zigzag of
pinning centers. The strongest pinning corresponds to the maximum
pinning sphere size, which is $R=3.0\xi$. In this case the
depinning transition occurs for $\theta_c=81^\circ$, and figure
\ref{figure5} shows that this maximum vortex line stretch is
$54\%$, thus  below the maximum limit of $61,8\%$ reached for
$90^\circ$, represented in figure \ref{figure5} through a dashed line.

\begin{figure*}[!t]
    \centering
    \subfigure[]
    {
    \begin{minipage}[t]{0.31\textwidth}
    \centering
    \includegraphics[width=\linewidth]{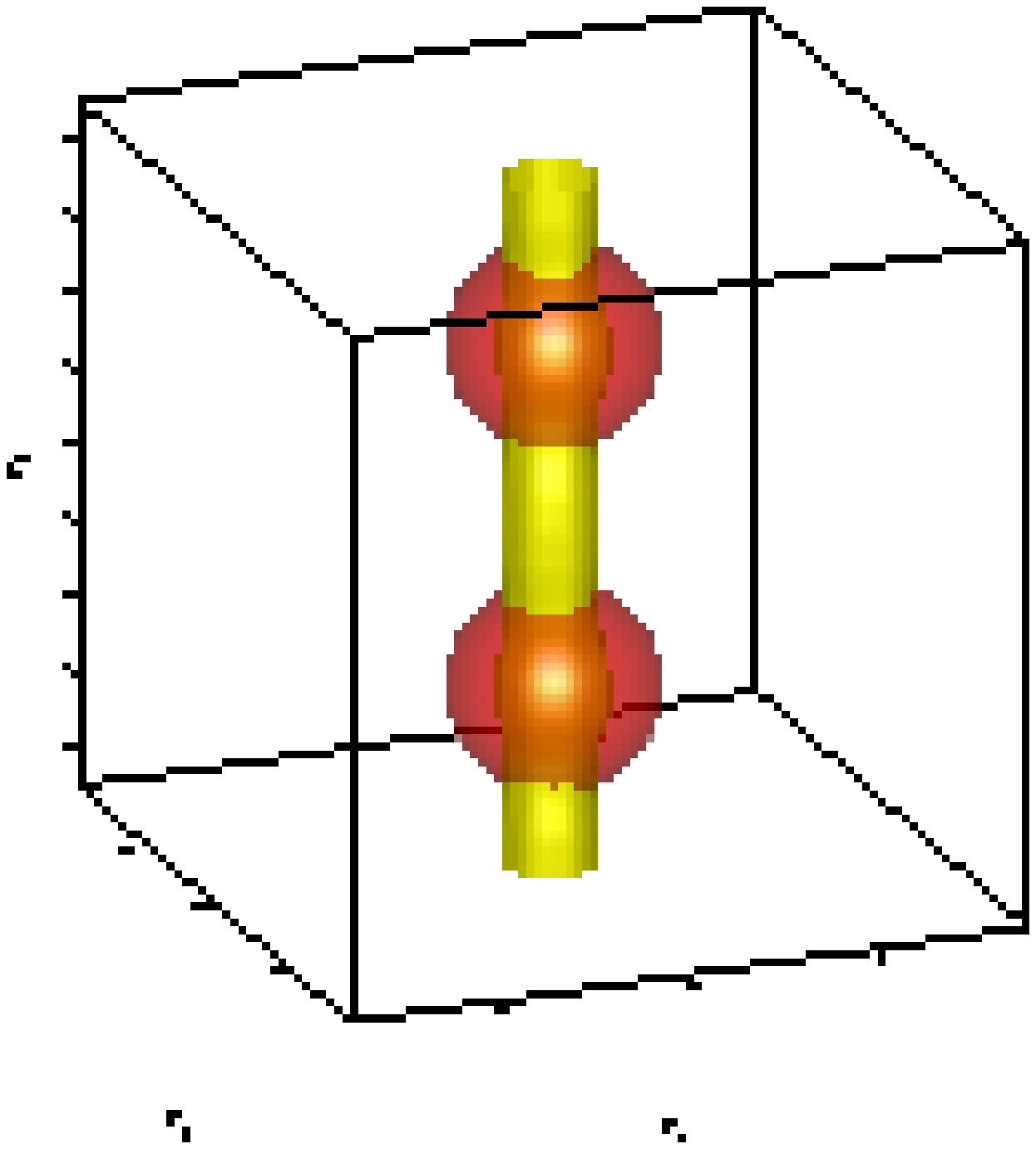}
    \label{dtn0R18}
    \end{minipage}
    }
    %\hfill
    \subfigure[]
    {
    \begin{minipage}[t]{0.31\textwidth}
    \centering
    \includegraphics[width=\linewidth]{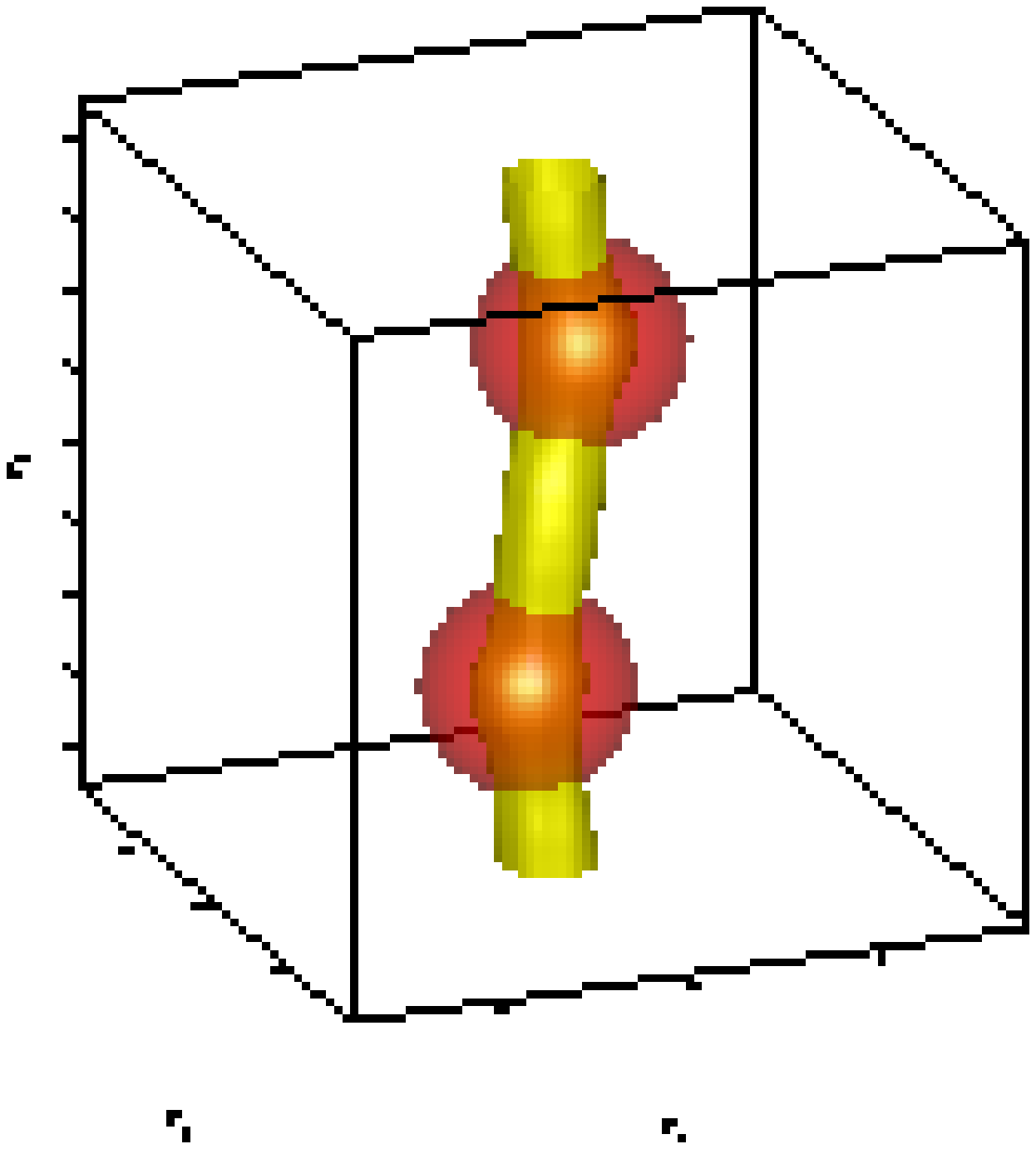}
    \label{dtn3R18}
    \end{minipage}
    }
    \subfigure[]
    {
    \begin{minipage}[t]{0.31\textwidth}
    \centering
    \includegraphics[width=\linewidth]{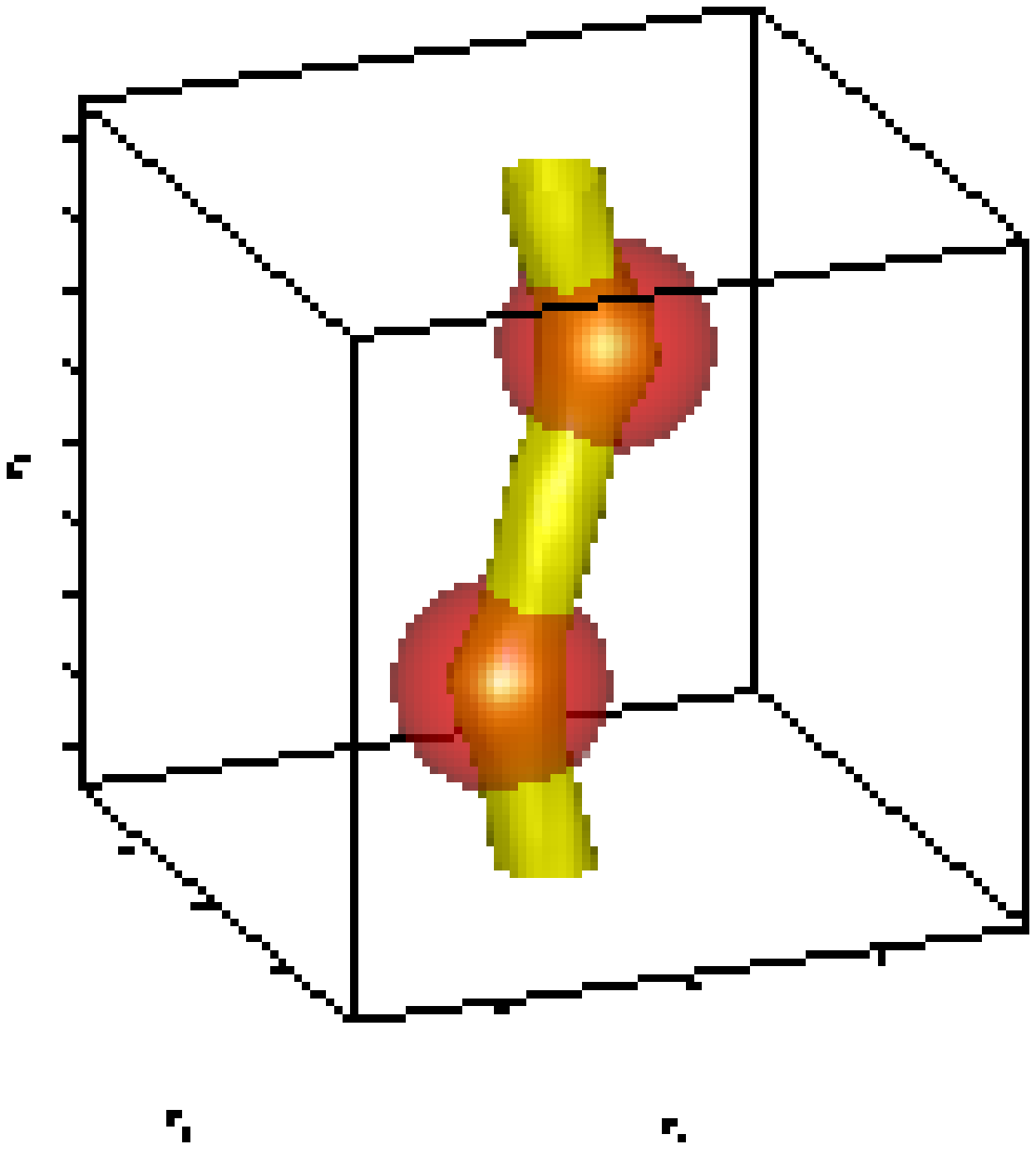}
    \label{dtn6R18}
    \end{minipage}
    }
    %\hfill
    \subfigure[]
    {
    \begin{minipage}[t]{0.31\textwidth}
    \centering
    \includegraphics[width=\linewidth]{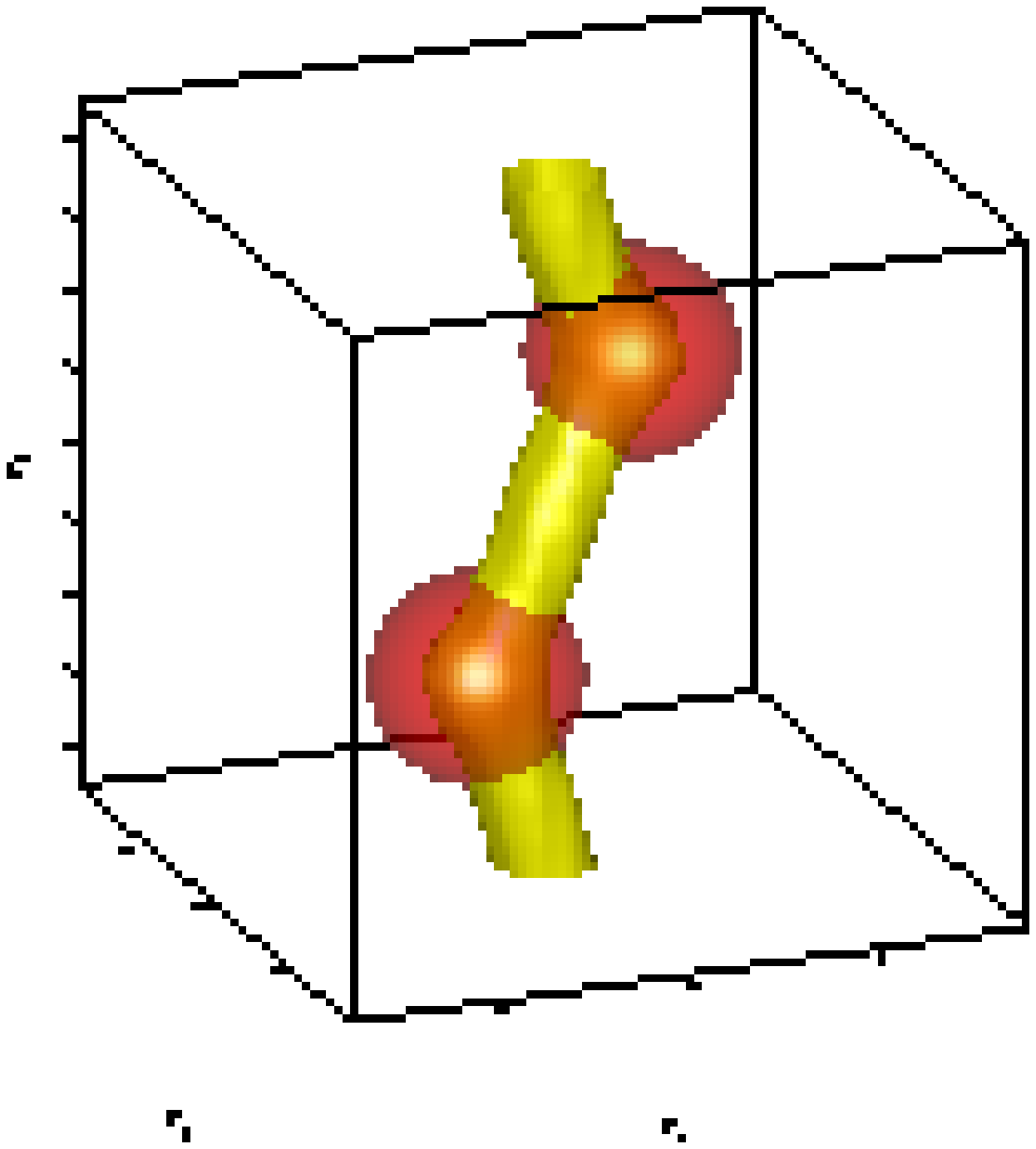}
    \label{dtn9R18}
    \end{minipage}
    }
    %\hfill
    \subfigure[]
    {
    \begin{minipage}[t]{0.31\textwidth}
    \centering
    \includegraphics[width=\linewidth]{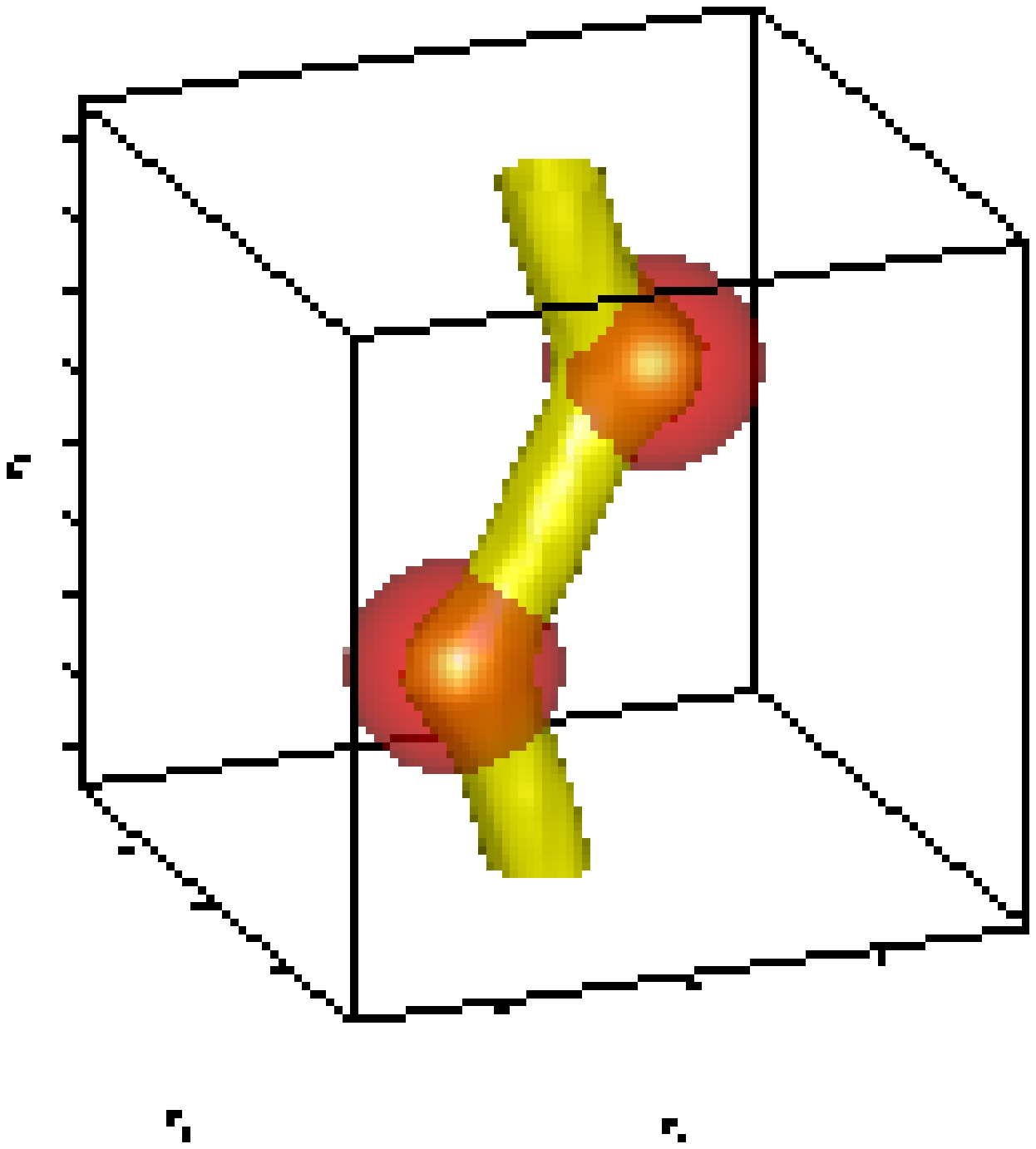}
    \label{dtn12R18}
    \end{minipage}
    }
    %\hfill
    \subfigure[]
    {
    \begin{minipage}[t]{0.31\textwidth}
    \centering
    \includegraphics[width=\linewidth]{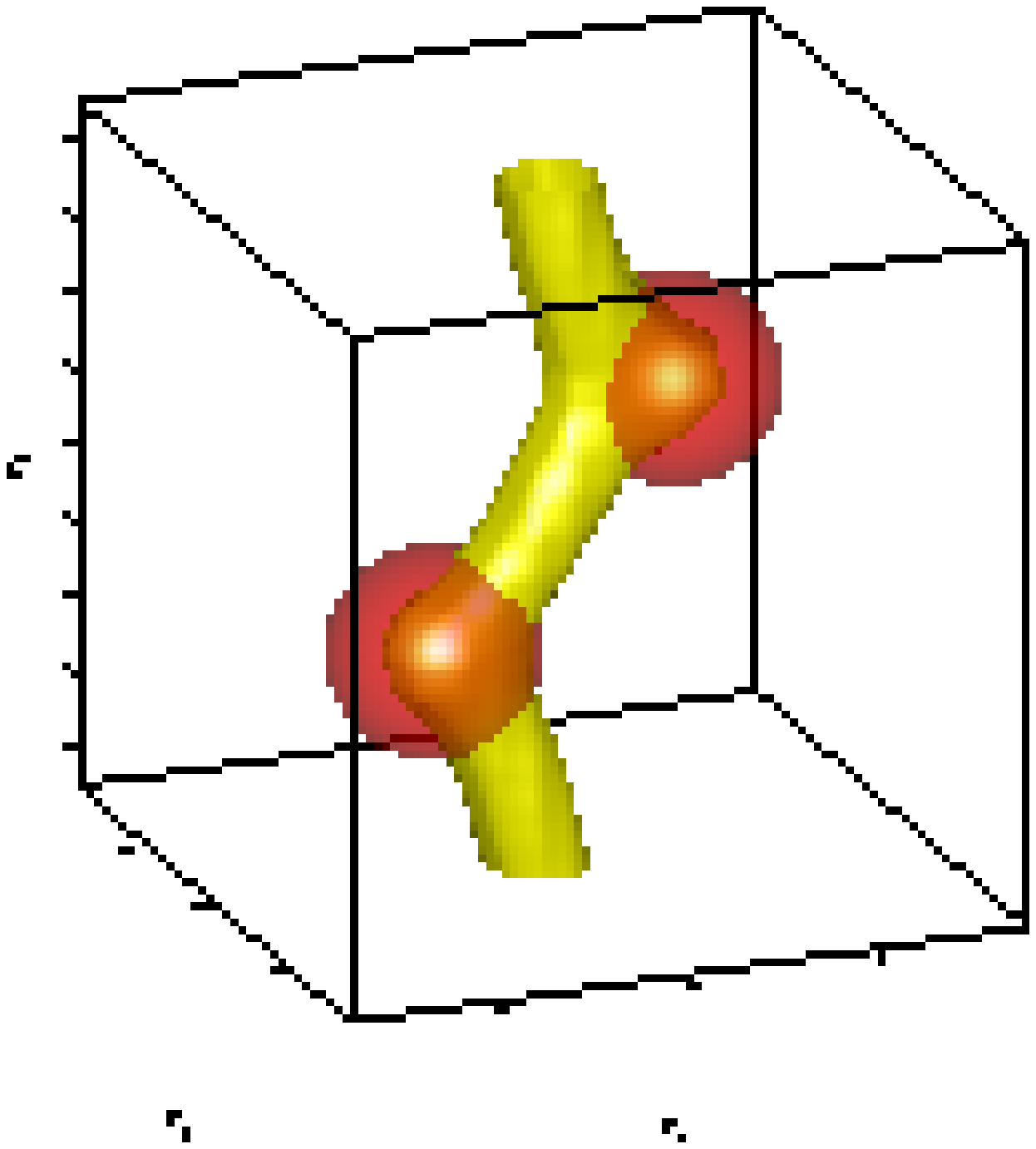}
    \label{dtn15R18}
    \end{minipage}
    }
    %\hfill
    \subfigure[]
    {
    \begin{minipage}[t]{0.31\textwidth}
    \centering
    \includegraphics[width=\linewidth]{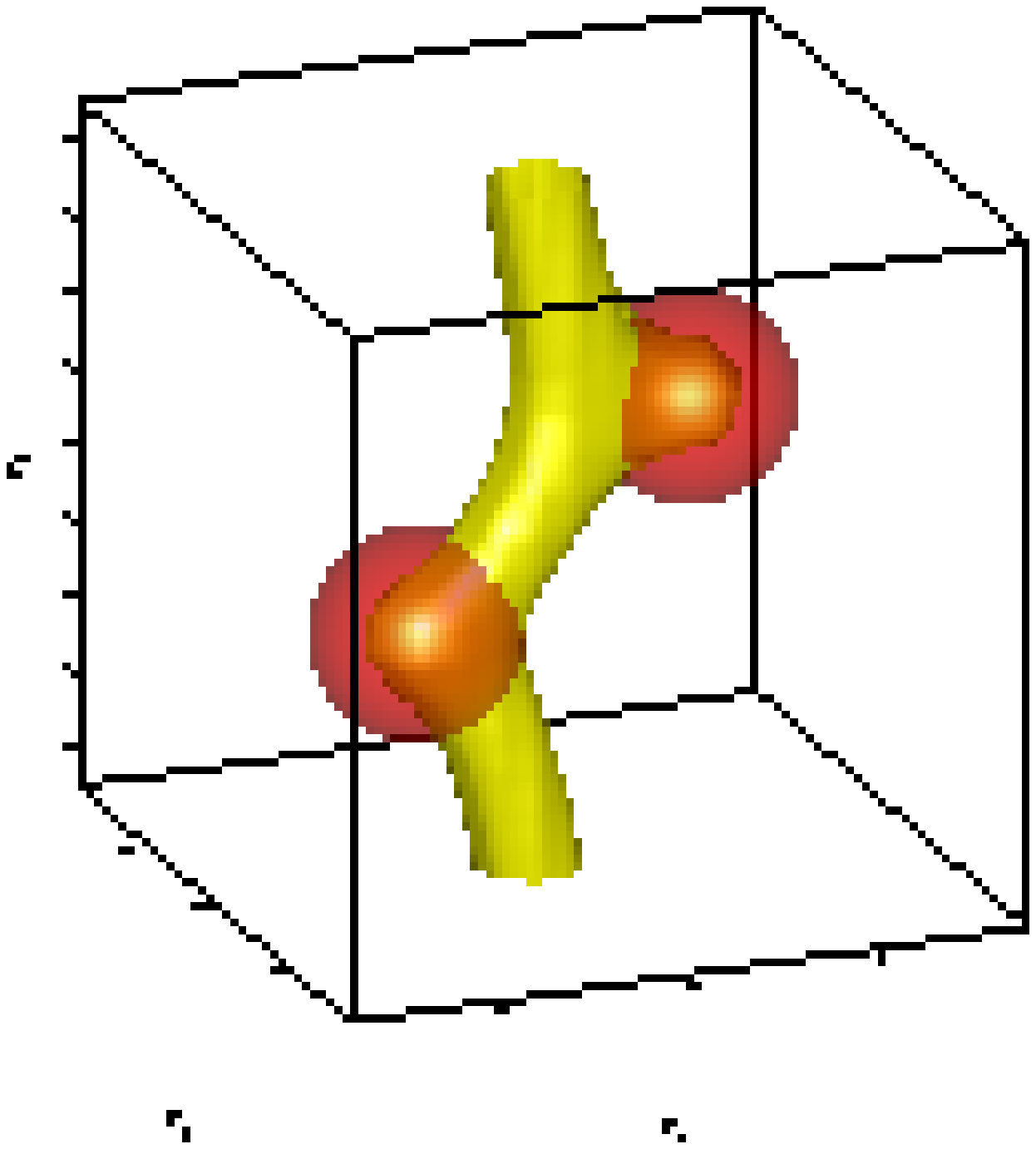}
    \label{dtn18R18}
    \end{minipage}
    }
    %\hfill
    \subfigure[]
    {
    \begin{minipage}[t]{0.31\textwidth}
    \centering
    \includegraphics[width=\linewidth]{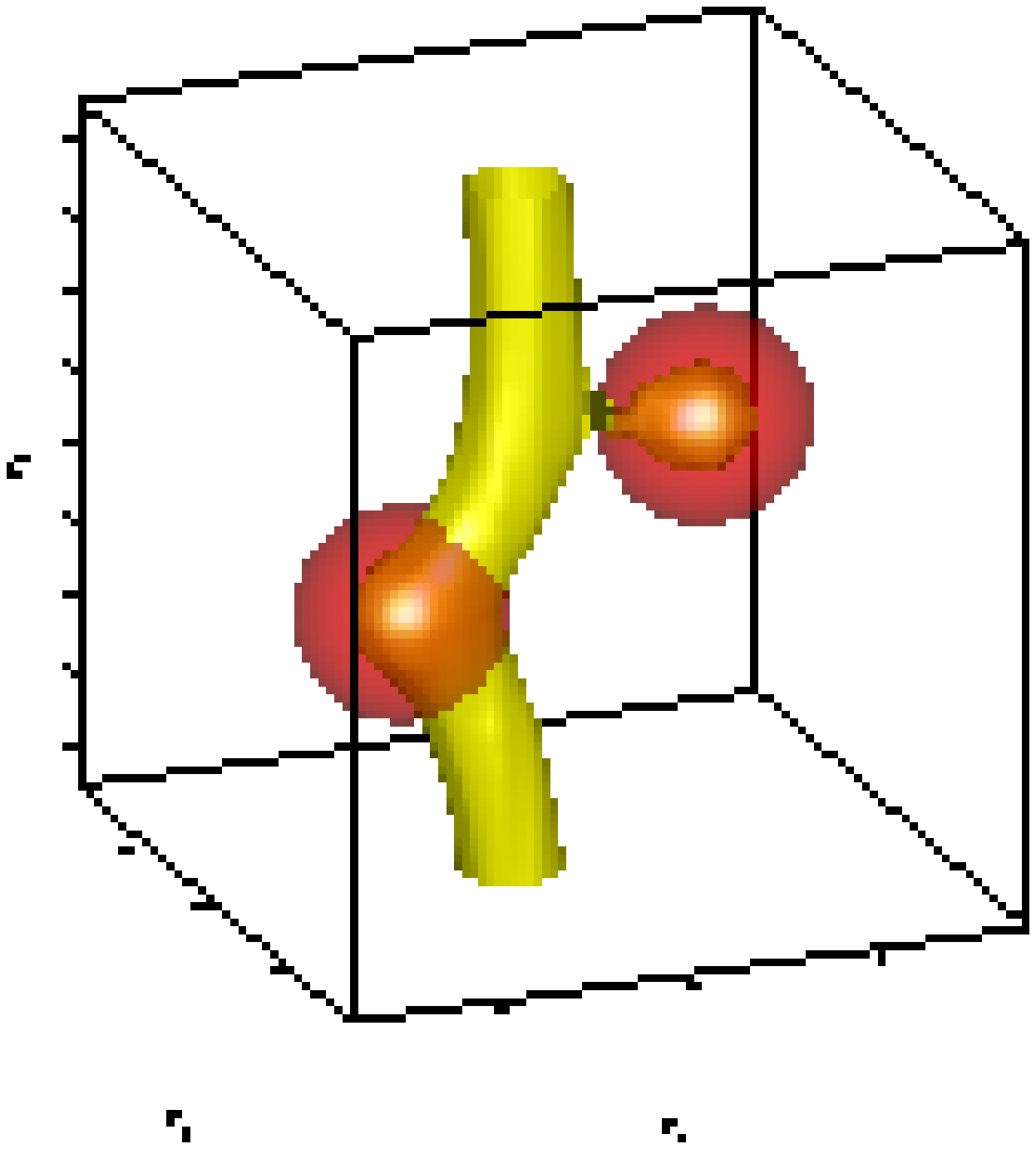}
    \label{dtn21R18}
    \end{minipage}
    }
    %\hfill
    \subfigure[]
    {
    \begin{minipage}[t]{0.31\textwidth}
    \centering
    \includegraphics[width=\linewidth]{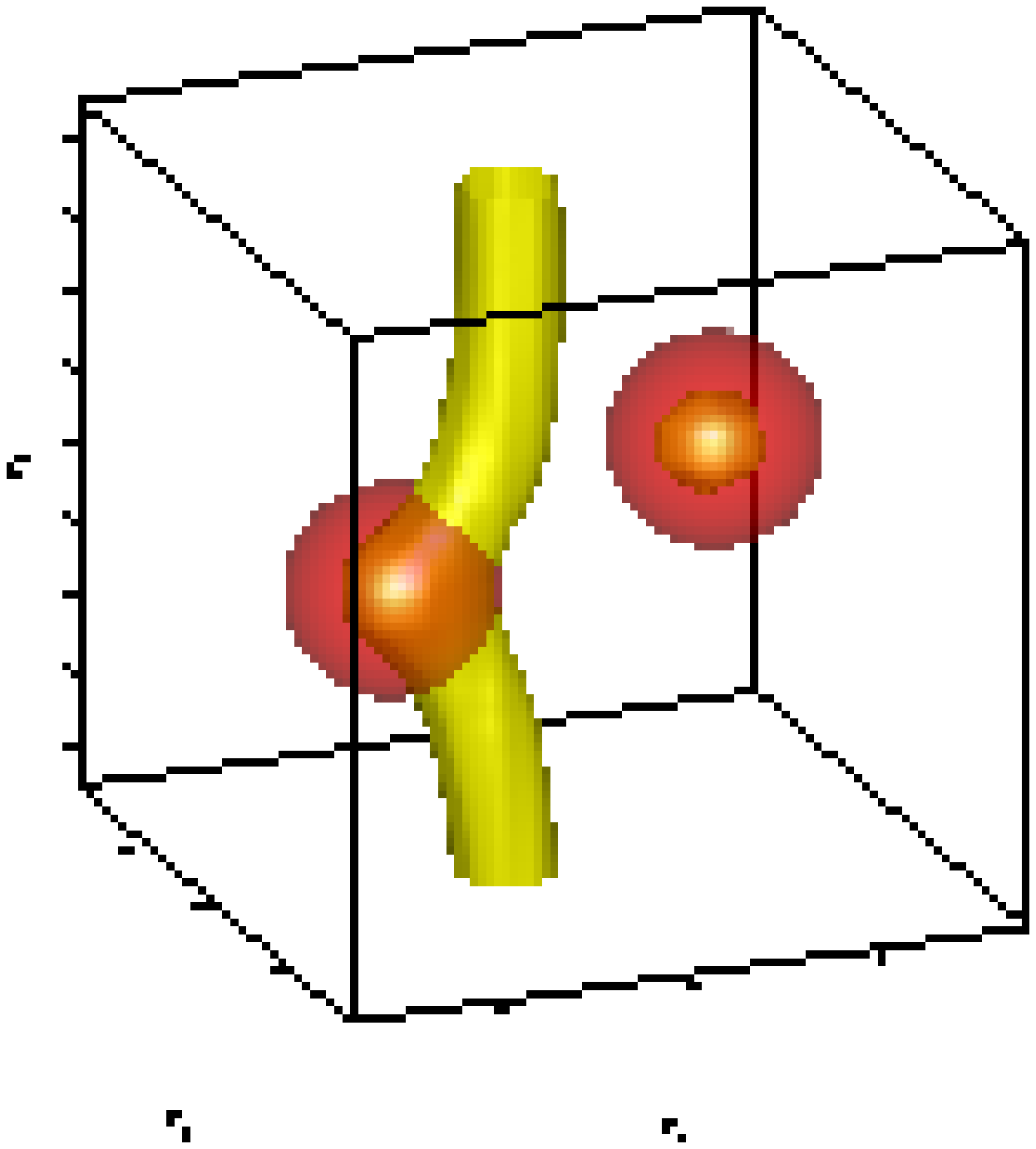}
    \label{dtn24R18}
    \end{minipage}
    }
    \caption{Visualization of the order parameter density $\left |\Delta\right |^2$ and of the pinning centers
    of radius $R=1.8\xi$. Extreme density values in the unit cell are $\left |\Delta\right |^2_{max}=0,9990$ and
    $\left |\Delta \right |^2_{min}=0,0034$. The density is visualized for  $\left |\Delta\right |^2_{iso}=0,2522$.
    The figures from \ref{dtn0R18} to \ref{dtn24R18} correspond to $\theta$ equal to 0$^\circ$, 9$^\circ$, 18$^\circ$,
    27$^\circ$, 36$^\circ$, 45$^\circ$, 54$^\circ$, 63$^\circ$ and 72$^\circ$ degree, respectively.}
     \label{framesR18}
\end{figure*}

\begin{figure*}[!t]
    \centering
    \subfigure[]
    {
    \begin{minipage}[t]{0.31\textwidth}
    \centering
    \includegraphics[width=\linewidth]{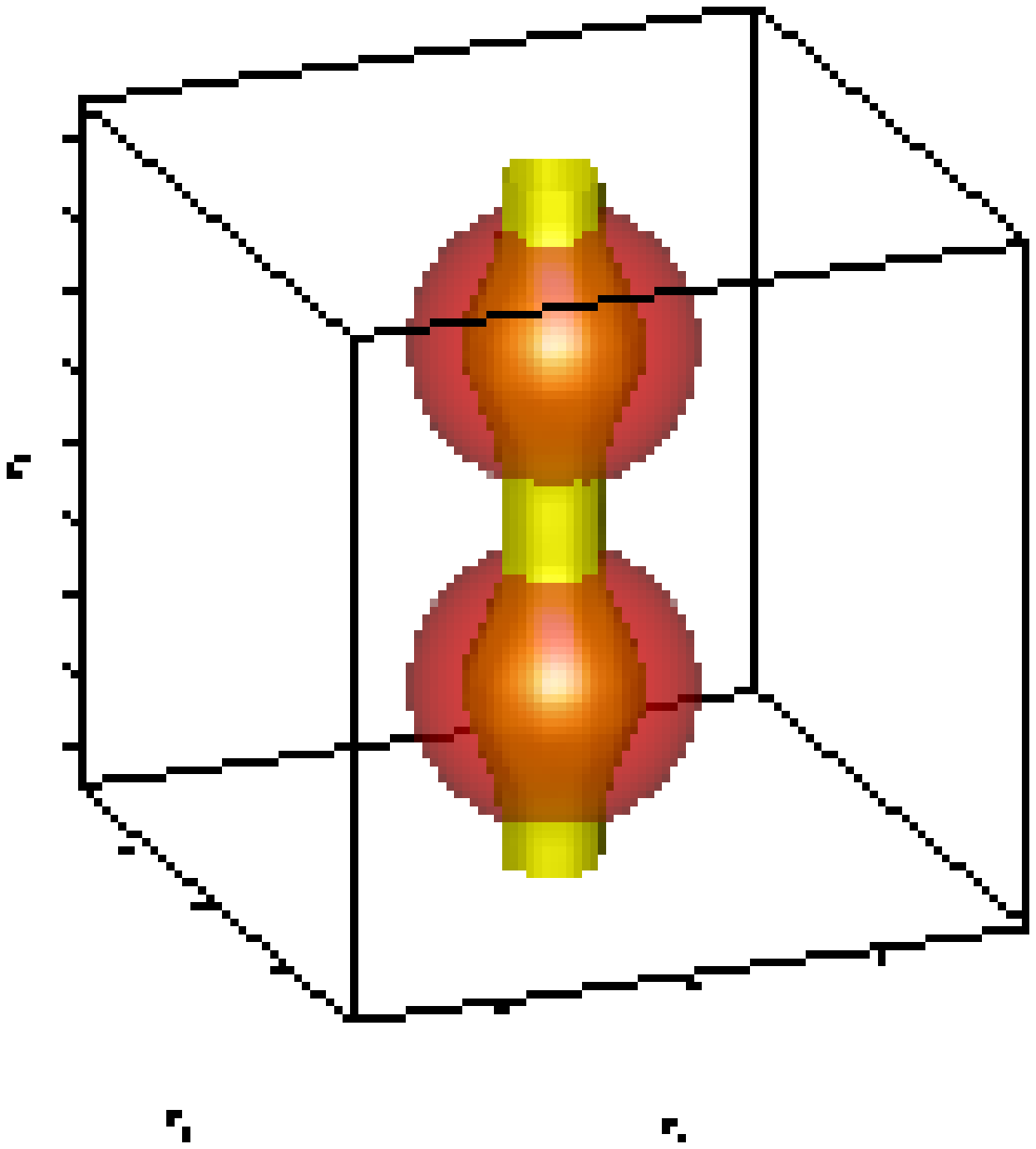}
    \label{dtn0R24}
    \end{minipage}
    }
    %\hfill
    \subfigure[]
    {
    \begin{minipage}[t]{0.31\textwidth}
    \centering
    \includegraphics[width=\linewidth]{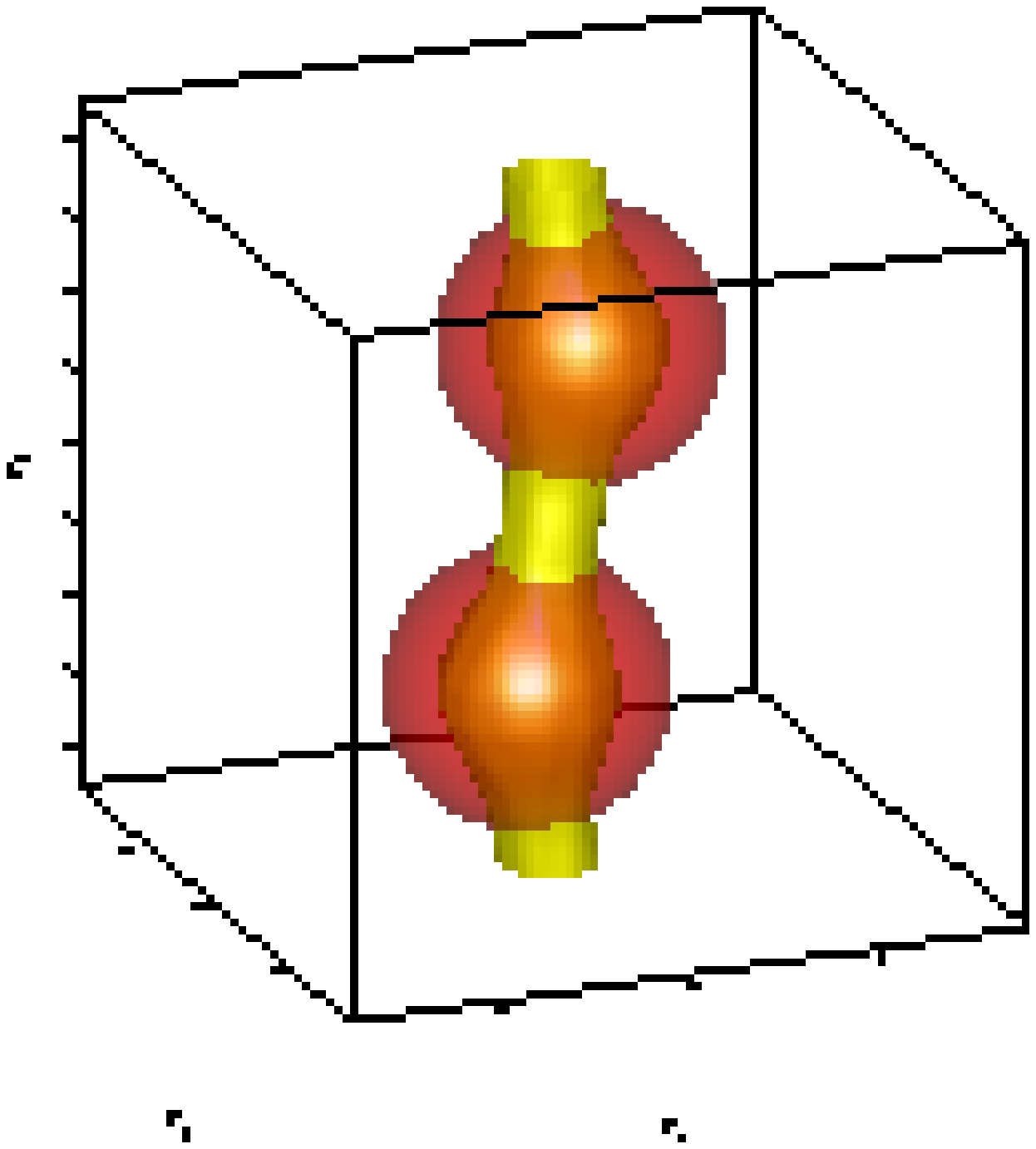}
    \label{dtn3R24}
    \end{minipage}
    }
    \subfigure[]
    {
    \begin{minipage}[t]{0.31\textwidth}
    \centering
    \includegraphics[width=\linewidth]{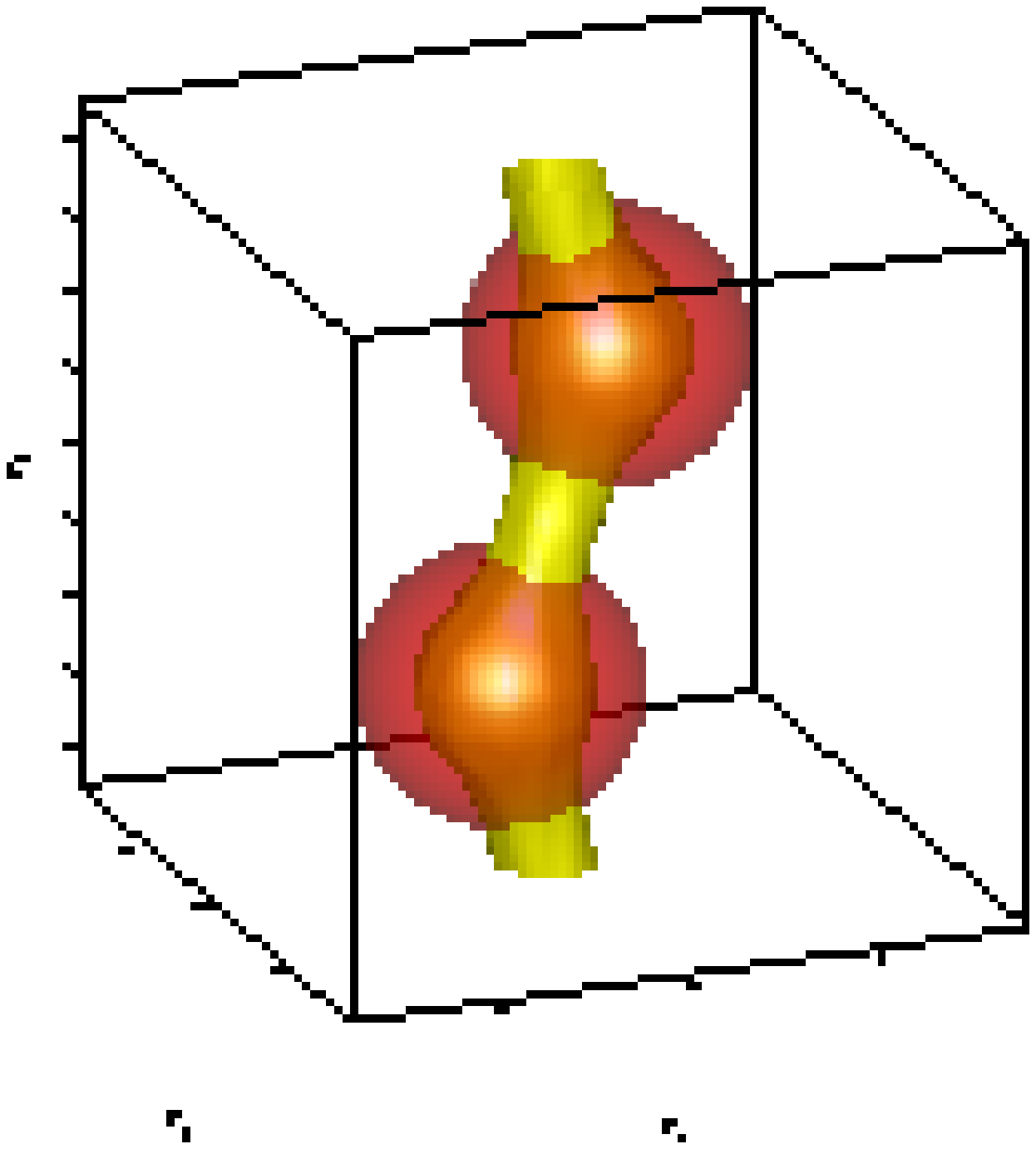}
    \label{dtn6R24}
    \end{minipage}
    }
    %\hfill
    \subfigure[]
    {
    \begin{minipage}[t]{0.31\textwidth}
    \centering
    \includegraphics[width=\linewidth]{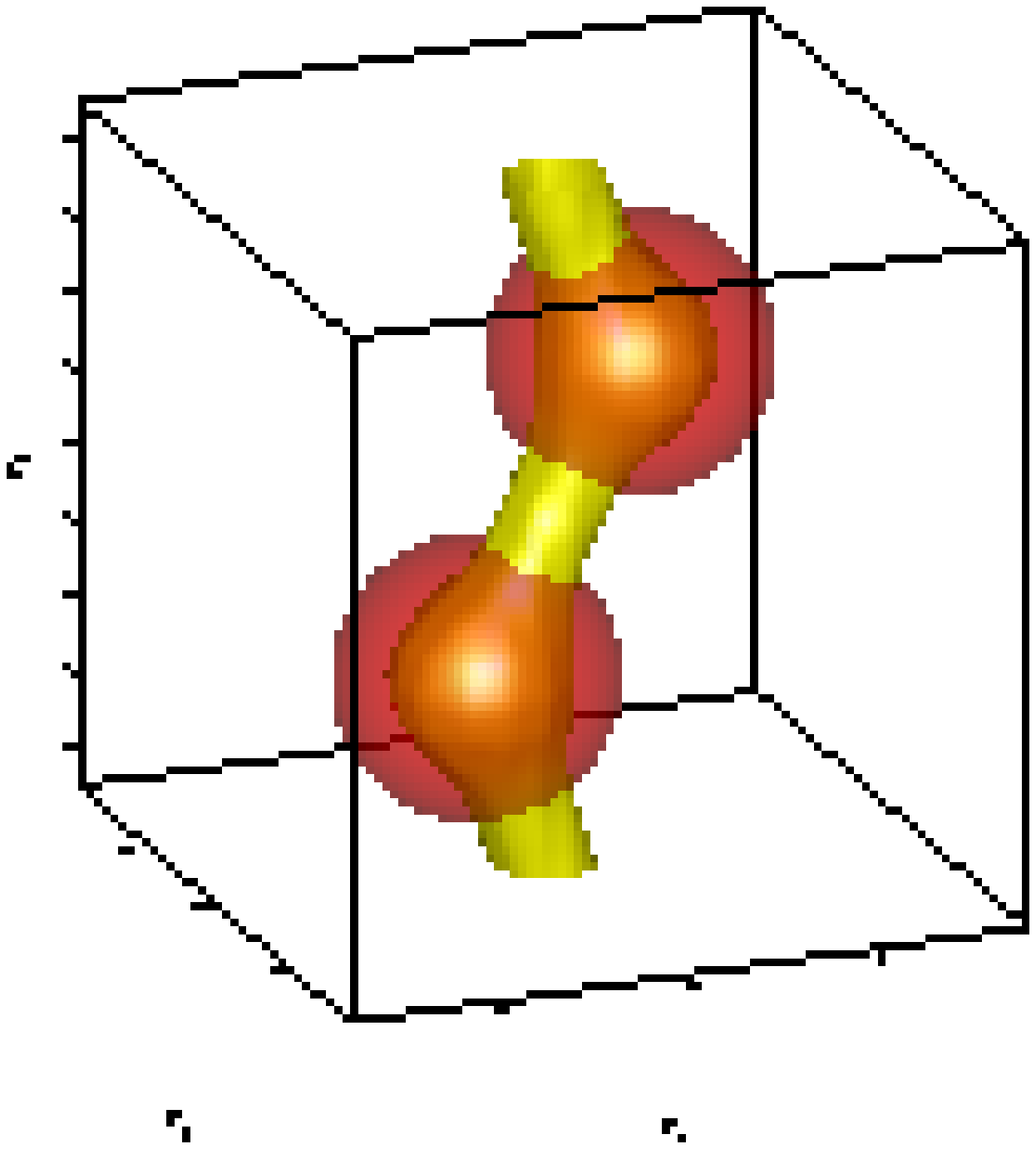}
    \label{dtn9R24}
    \end{minipage}
    }
    %\hfill
    \subfigure[]
    {
    \begin{minipage}[t]{0.31\textwidth}
    \centering
    \includegraphics[width=\linewidth]{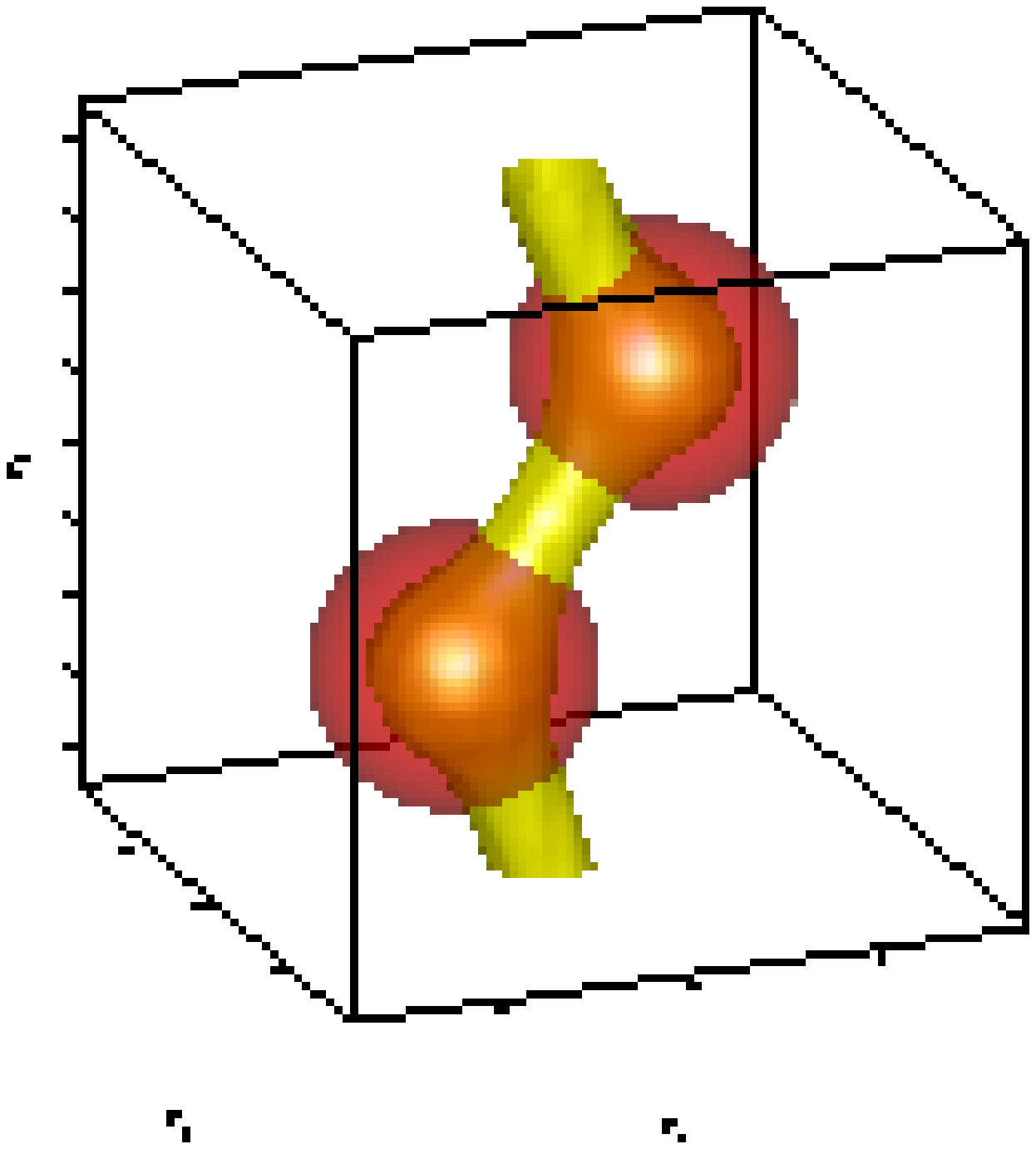}
    \label{dtn12R24}
    \end{minipage}
    }
    %\hfill
    \subfigure[]
    {
    \begin{minipage}[t]{0.31\textwidth}
    \centering
    \includegraphics[width=\linewidth]{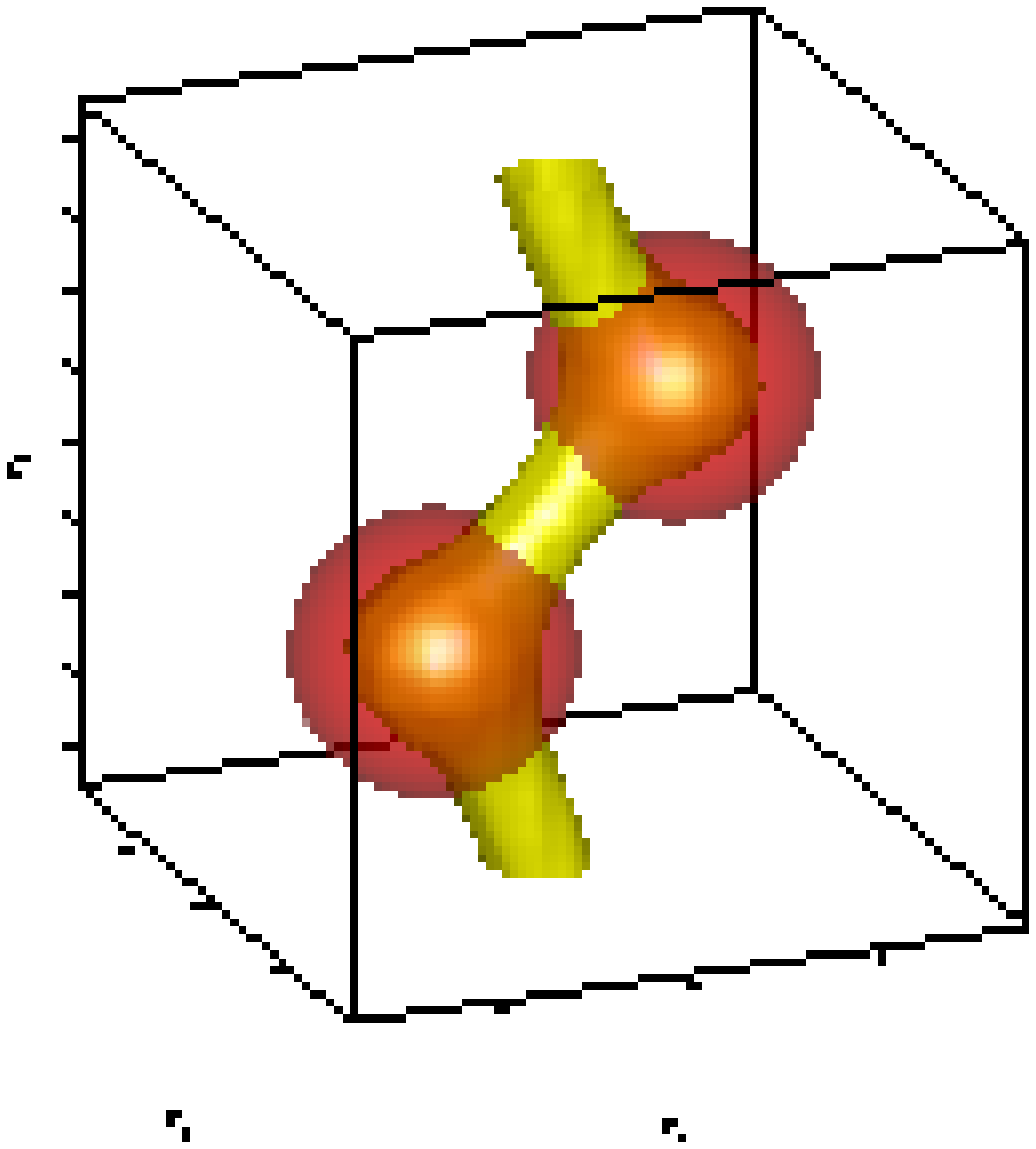}
    \label{dtn15R24}
    \end{minipage}
    }
    %\hfill
    \subfigure[]
    {
    \begin{minipage}[t]{0.31\textwidth}
    \centering
    \includegraphics[width=\linewidth]{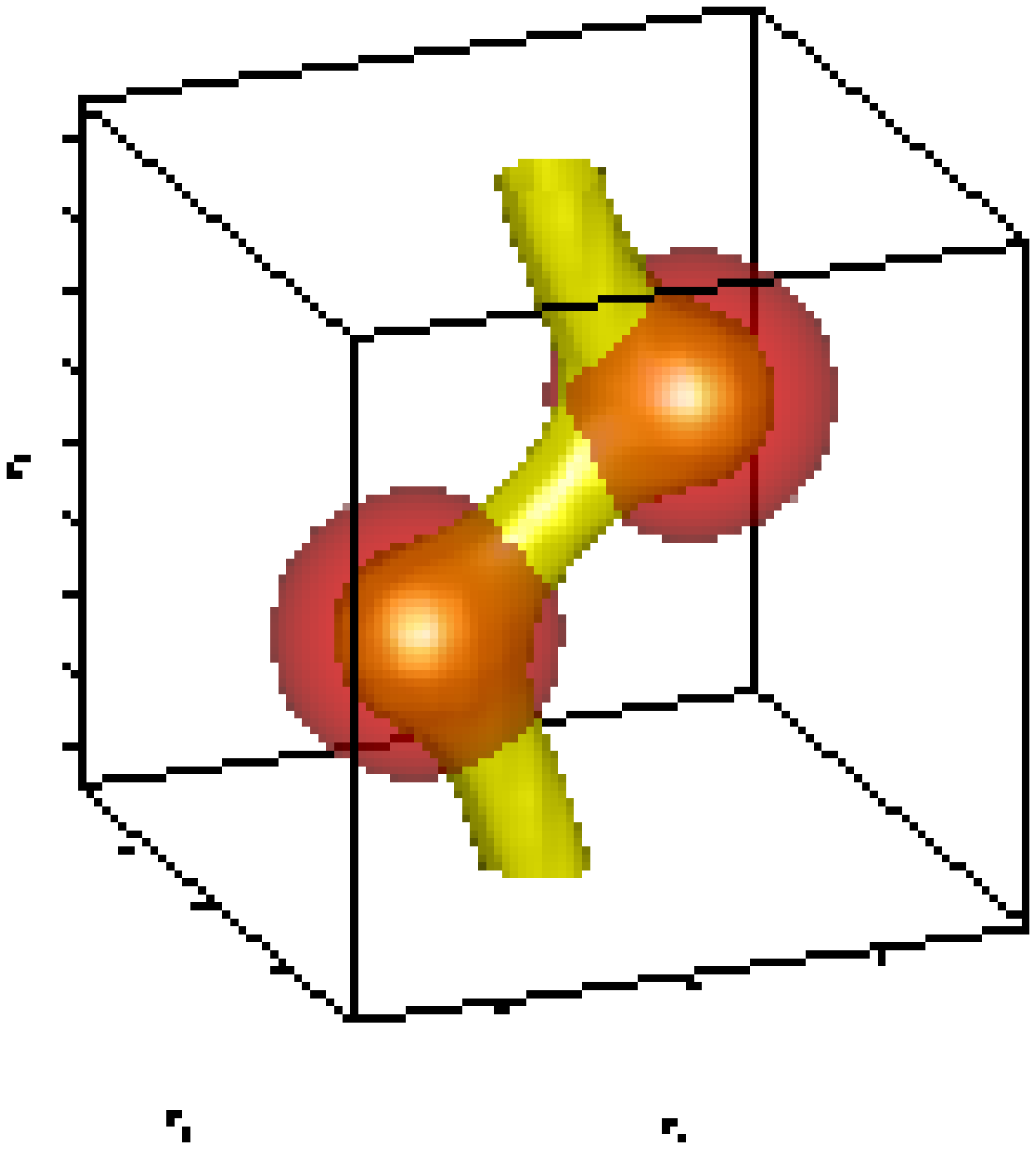}
    \label{dtn18R24}
    \end{minipage}
    }
    %\hfill
    \subfigure[]
    {
    \begin{minipage}[t]{0.31\textwidth}
    \centering
    \includegraphics[width=\linewidth]{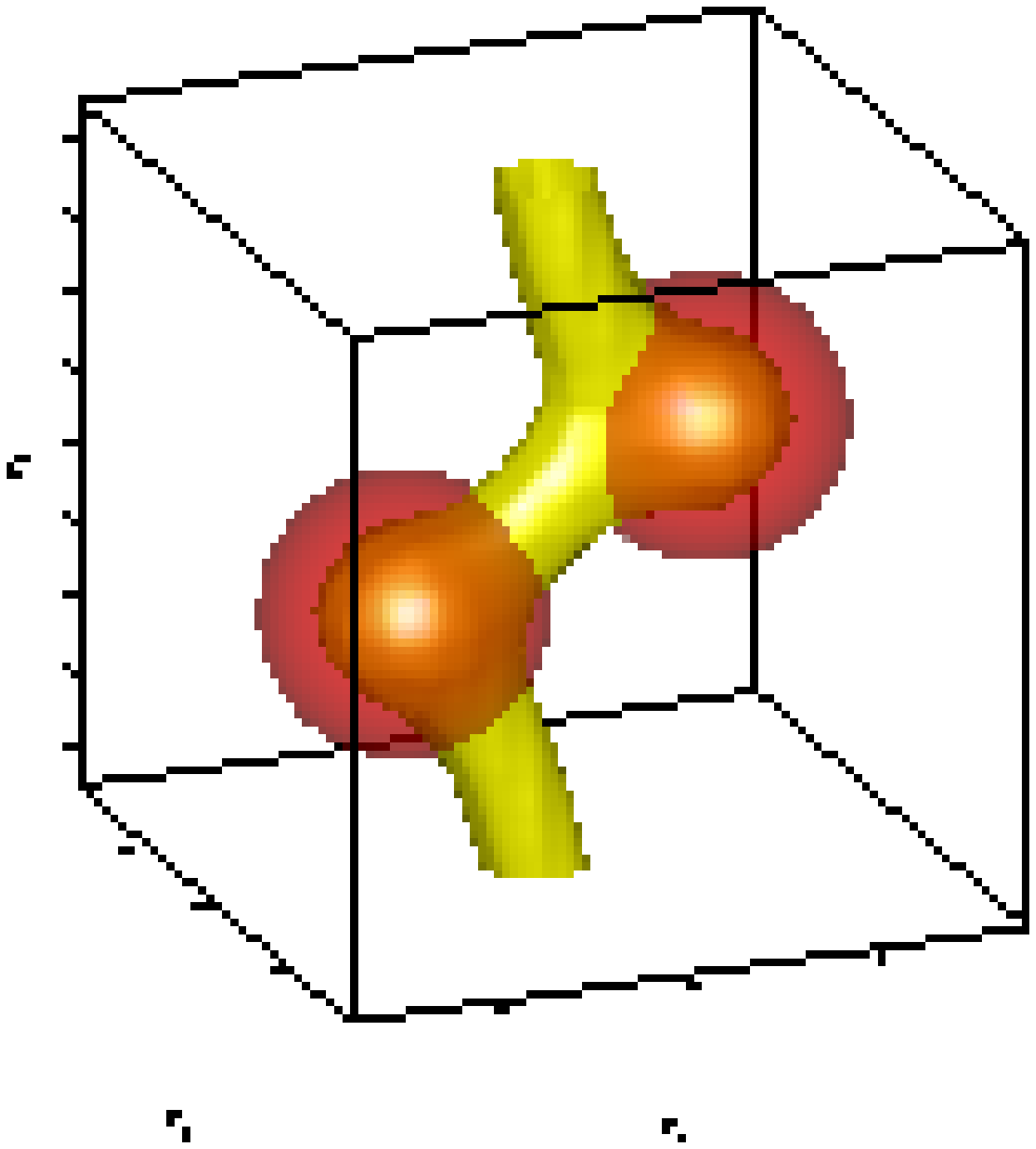}
    \label{dtn21R24}
    \end{minipage}
    }
    %\hfill
    \subfigure[]
    {
    \begin{minipage}[t]{0.31\textwidth}
    \centering
    \includegraphics[width=\linewidth]{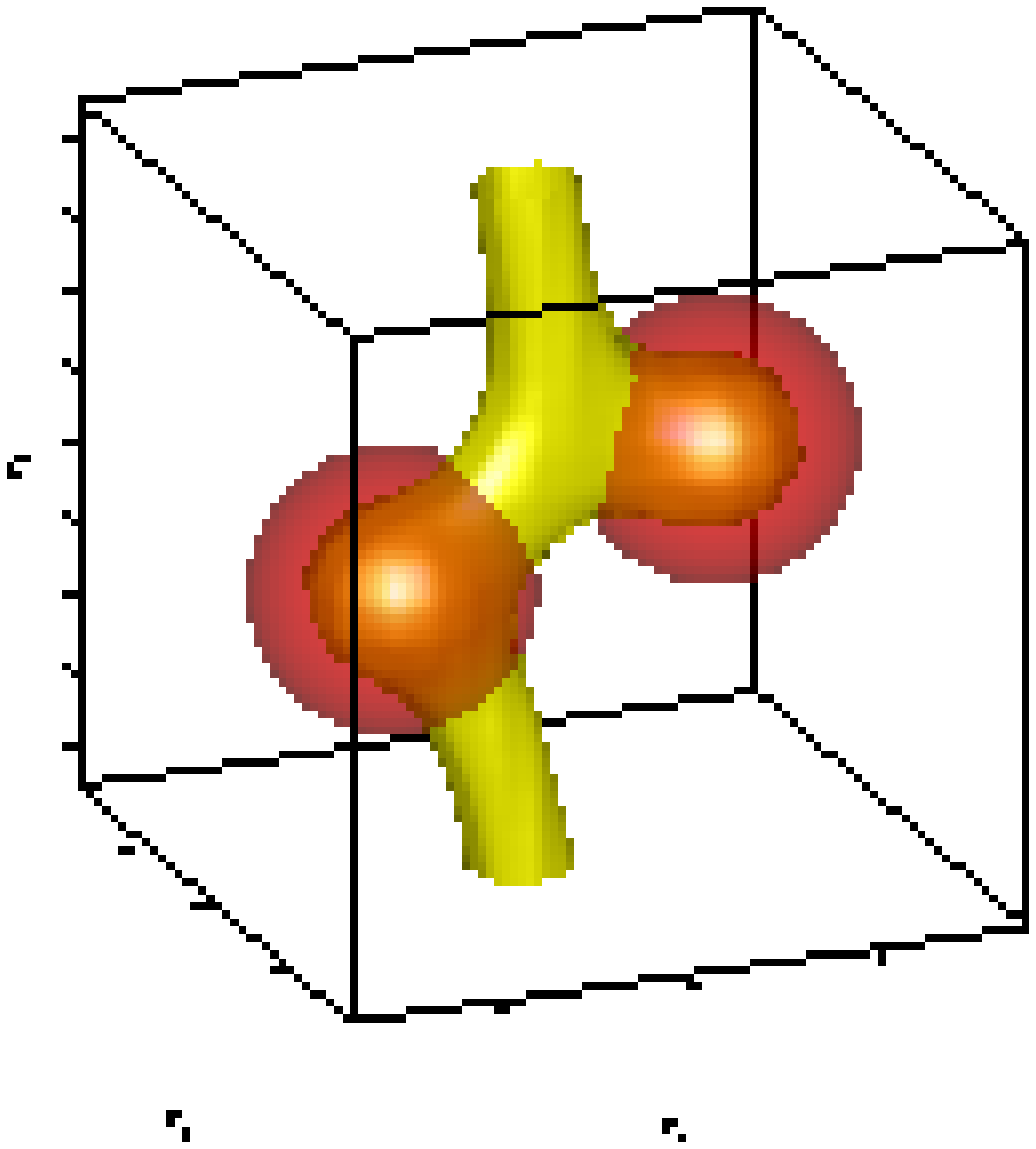}
    \label{dtn24R24}
    \end{minipage}
    }
    \caption{Visualization of the order parameter density $\left |\Delta\right |^2$ and of the pinning centers
    of radius $R=2.4\xi$. Extreme density values in the unit cell are $\left |\Delta\right |^2_{max}=0,9990$
    and $\left |\Delta\right |^2_{min}=0,0035$.
    The density is visualized for $\left |\Delta\right |^2_{iso}=0,2507$.
    The  figures from \ref{dtn0R24} to \ref{dtn24R24} correspond to $\theta$ equal
     to 0$^\circ$, 9$^\circ$, 18$^\circ$, 27$^\circ$, 36$^\circ$, 45$^\circ$, 54$^\circ$, 63$^\circ$
     and 72$^\circ$ degree, respectively.}
    \label{framesR24}
\end{figure*}

The present results are best understood through figures
\ref{framesR18} and \ref{framesR24} which show semi-transparent
views of the vortex and of the pinning sphere surfaces\cite{tonhomepage}.

Figures \ref{dtn0R18} and \ref{dtn0R24} show that for $R
> \xi$ the pinning sphere volume is partially outside the vortex
core. As previously discussed this extra insulating volume costs
energy causing the free energy of the superconductor to approach
the normal state, which means to increase with $R$, as shown in
figure \ref{figure3a}. Figures \ref{framesR18} and \ref{framesR24}
show three-dimensional plots of the normalized density
$|\Delta|^2$ for radii $R=1.8\xi$ and $R=2.4\xi$, respectively,
taken at angles $0^\circ$, $9^\circ$, $18^\circ$, $27^\circ$,
$36^\circ$, $45^\circ$, $54^\circ$, $63^\circ$, and $72^\circ$.
%---------------------figure  #5------------------------------------------------------
%---------------------figure  #5------------------------------------------------------
Each tridimensional figure shows an iso-surface of constant
density inside the unit cell, taken here as a fraction of the
maximum value of the normalized density,
$|\Delta|_{iso}^2=(|3\Delta|_{max}^2+|\Delta|_{min}^2)/4$. It also
shows the pinning sphere surfaces and this  provides a way to see
in these plots the length scales of $1.8\xi$ and $2.4\xi$,
respectively. Notice that above the depinning transition there is
an iso-surface sphere, $|\Delta|_{iso}^2$, inside the pinning
spheres. The order parameter does not drop abruptly to zero inside
the pinning spheres because the functions $\tau_i$, taken in our
calculations, are smooth versions of the Heaviside function,
defined as $\tau = 1$ outside the spheres and $\tau = 0$ inside
them. Abrupt changes within a distance smaller than two neighbor
mesh points can lead to numerical instability, and for this reason
some degree of smoothness is necessary because of our coarse-grain
treatment of this problem. Figure \ref{framesR18} shows the
behavior of the vortex line inside the unit cell for several
angular arrangements of the two pinning spheres with $R=1.8\xi$,
ranging from figure \ref{dtn0R18} until the depinning transition
in figure \ref{dtn18R18}. Pinning of the vortex by the top sphere
becomes increasingly more difficult as $\theta$ increases
resulting that figures \ref{dtn21R18} and \ref{dtn24R18} describe
configurations above the depinning transition. Figure
\ref{framesR24} shows a similar sequence of angular arrangements
with the depinning transition corresponding to figure
\ref{dtn24R24}. For both radii, as well as for any other one, we
find that the depinning always occurs from the top sphere and
never from the bottom one. This apparent breaking of symmetry is a
consequence of the way the numerical procedure is carried here.
All the sub-figures of figure \ref{framesR18} were independently
obtained from each other  in our numerical simulations, and the
same is true for the sub-figures of  figure \ref{framesR24}. Each
one of them results from an initial arbitrary configuration of the
order parameter that evolves as the temperature of the Simulating
Annealed method is lowered and the final configuration is reached
within some convergence criteria. The reason for this breaking of
symmetry is the way that the order parameter configuration is
updated during the minimization procedure. The order parameter is
updated in sweeps of the mesh points of the unit cell that start
from the bottom and end in the top. For this reason the vortex
line remains pinned to the bottom sphere.
%
%
%........................................................................
\section{Conclusions}
\label{sec:con}
The length a vortex line inside a superconductor with pinning
centers in case of no thermal fluctuations, is not just determined
by the direction of the applied field and the geometry of  the
sample. The vortex line is subjected to the two conflictual
demands of pinning by as many defects as possible, and its
consequent increase in self-energy. We have studied here this
problem in the context of a simple model with just one kind of
pinning center, insulating spheres of coherence length size
radius, forming a zigzag near the vortex line. The zigzags are
periodically arranged and form  a lattice, which is simply
described by a cubic unit cell with two pinning centers inside,
whose center of the segment that connects them coincides with the
center of the cube. This segment is free to rotate around its
center producing for each angle a different zigzag path, whose
pinning centers interact with the vortex line that pierces a pile
of unit cell along the z-axis. The simplest possible zigzag
arrangement is the straight line of insulating spheres of equal
radii intercalated by two segments that pins the vortex line along
the z-axis. As the pinning spheres inside the unit cell rotate a
different zigzag of defects is produced, resulting for each
rotation a vortex line more and more deviated from the straight
line. The balance between the two competing effects, that is,
defect trapping and self energy, changes with angle to the point
that a depinning transition occurs and the vortex  becomes  nearly
straight again. This maximum stretch of the vortex line depends on
the pinning strength, here associated to the radius of the pinning
sphere. We have numerically determined this depinning transition
for a special pinning center lattice, with density, $2/L^3$,
$L=12.0\xi$, and several pinning strength, represented by radii,
ranging from $R=1.2\xi$ to $R=3.0\xi$. We find that for this
lattice the maximum length increase of the vortex line is  54\%
bigger than the straight line.
%
%........................................................
\section{Acknowledgments}
\label{Acknowledgments}
Research supported in part by Instituto do Mil\^enio de
Nano-Ci\^encias, CNPq, and FAPERJ (Brazil).
%\pagestyle{plain}
%
%..............................................................
\bibliographystyle{unsrt}
\bibliography{EPJB}
%
%---------------------figure  #6------------------------------------------------------
%---------------------figure  #6------------------------------------------------------
%
%
%---------------------figure  #7------------------------------------------------------
%---------------------figure  #7------------------------------------------------------
\end{document}